RESEARCH ARTICLE

# Windows Instant Messaging App Forensics: Facebook and Skype as Case Studies

Teing Yee Yang[1], Ali Dehghantanha[2], Kim-Kwang Raymond Choo[3]*, Zaiton Muda[1]

1 Department of Computer Science, Faculty of Computer Science and Information Technology, Universiti Putra Malaysia, UPM Serdang, Selangor, Malaysia, 2 The School of Computing, Science & Engineering, Newton Building, University of Salford, Salford, Greater Manchester, United Kingdom, 3 Information Assurance Research Group, University of South Australia, Adelaide, South Australia, Australia

* raymond.choo@fulbrightmail.org

## Abstract

Instant messaging (IM) has changed the way people communicate with each other. However, the interactive and instant nature of these applications (apps) made them an attractive choice for malicious cyber activities such as phishing. The forensic examination of IM apps for modern Windows 8.1 (or later) has been largely unexplored, as the platform is relatively new. In this paper, we seek to determine the data remnants from the use of two popular Windows Store application software for instant messaging, namely Facebook and Skype on a Windows 8.1 client machine. This research contributes to an in-depth understanding of the types of terrestrial artefacts that are likely to remain after the use of instant messaging services and application software on a contemporary Windows operating system. Potential artefacts detected during the research include data relating to the installation or uninstallation of the instant messaging application software, log-in and log-off information, contact lists, conversations, and transferred files.

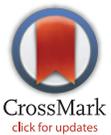

**Data Availability Statement:** All relevant data are within the paper.

**Funding:** These authors have no support or funding to report.

**Competing Interests:** The authors have declared that no competing interests exist.

## 1. Introduction

Instant messaging (IM) is popular with both traditional computing device users (i.e., personal computers and laptops) and mobile device users by allowing them to exchange information with peers in real time using text messaging, voice messaging, and file sharing. According to the report of Radicati Group [1], the number of worldwide IM accounts (with the exception of mobile messaging) in 2015 amounted to over 3.2 billion which is expected to rise above 3.8 billion by the end of 2019.

Similar to other popular consumer technologies, IM services have also been exploited to commit frauds and scams [2–4], disseminate malware [5], groom children online with the purpose of sexual exploitation [6–9] etc. The chat logs can provide a great deal of information of evidential value to investigators [10, 11], which may often comprise a suspect's physical location, true identity, transactional information, incriminating conversations, and other person information i.e., email address and bank account number [12].

Due to the increased user privacy requirements [13] and demands for data redundancy, it is increasingly challenging to collect evidential data from the IM service provider (ISP). The data are often protected by proprietary protocols, encryption, etc., making forensic practitioners virtually impossible to collect meaningful information from external network [14]. Moreover, collecting data from a multi-tenancy environment may breach the data privacy policies of the ISPs [15]. Even if the artefacts could be identified, the challenges are compounded by cross-jurisdictional investigations that may prohibit cross-border transfer of information [16–18]. In the worst-case scenario, the ISPs may not even log the incriminating conversations to reduce traffic to the messaging servers [19].

Depending on the IM application in use, the client device can often provide potential for alternative methods for recovery of the IM artefacts [20–22]. In addition to addressing the possible issues in relation to evidence acquisition from the ISPs, the terrestrial artefacts can be useful in establishing whether a suspect has a direct connection to a crime, as the suspect may claim he/she is a victim of identity theft otherwise. While a practitioner should be cognisant of techniques of digital forensics, it is just as important to maintain an up-to-date understanding of the potential artefacts that are recoverable from different types of IM products. Hence, in this paper, we seek to identify potential terrestrial artefacts that may remain after the use of the popular Facebook and Skype Windows Store application software (henceforth the Store app) on a Windows 8.1 client machine. Similar to the approaches of Quick and Choo [23–25], we attempt to answer the following questions in this research:

1. What data remains on a Windows 8.1 device and their locations on a hard drive after a user has used Facebook app version 1.4.0.9 and Skype app version 3.1.0.1007.

2. What data remains in Random Access Memory (RAM) after a user has used the above IM services or apps on a Windows 8.1 device?

3. What data can be seen in network traffic?

Findings from this research will contribute to the forensic community's understanding of the types of terrestrial artefacts that are likely to remain after the use of IM services and apps on devices running the newer Windows operating system.

The structure of this paper is as follows. Section 2 discusses the background and related work. Section 3 outlines the research methodology and experiment environment and setup. In Sections 4 to 6, we present and discuss the findings from the IM apps. We then conclude the paper and outline potential future research areas in the last section.

## 2. Literature Review

A Windows Store app (formerly known as Metro app) mimics the touch-screen-friendly mobile apps, while retaining the traditional mouse and keyboard inputs [26]. The installation is handled exclusively by the Windows Store, which bypasses the execution of executable files [27]. The Store apps are licensed to Microsoft account, giving the users the right to install a same app on up to eighty-one different Windows 8 (or newer) desktop clients under the same login [28]. The concept also enables the users to roam the app credentials (stored within the Credential Locker) between the corresponding devices [29].

The Store apps are predominantly built on Windows Runtime. In addition to offering the developers a multi-language programming environment, the architecture isolates the apps from the file system for security and stability [26]. The app itself is a package (.APPX file) that incorporates the app's code, resources, libraries, and a manifest up to a combined limit of 8GB [26]. Each Store app is represented by a package ID, which is often denoted by the package name followed by its build version, the target platform, and the alphanumeric publisher identification (ID). The installation and application folders can be generally located in *%Program*

Files%\WindowsApps\[Package ID] and %localappdata%\packages\[Package ID] respectively [30, 31].

The application data, correspond to the app states [26], are stored in three (3) categories: local, roaming, and temp states; each of which creates a subfolder in the application folder. The 'LocalState' folder holds device-specific data typically loaded to support the app functionality, such as temporary files and caches, recently viewed items, and other behavioural settings. The 'RoamingState' folder stores data shared between the same app running on multiple Windows devices under the same login. The data may include account configurations, favourites, game scores and progress, important URIs etc. Meanwhile, the 'TempState' folder houses data temporarily suspended or terminated from the memory for restoration purposes, such as page navigation history, unsaved form data etc. The application data persist throughout the lifetime of a Store app, with the exception of the temp data which may be subject to disk clean up [26].

The application cache/data can be stored using caching mechanisms like HTML5 local storage and IndexedDB (for Store apps written in HTML and JavaScript) as well as other third-party database options like SQLite [32]. In the absence of encryption mechanism, the data can aid in reconstruction of user events such as cloud storage [28], emails [30], web browsing history [33], conversations [34], and other user-specific events [35], depending on the Store app in use.

Instant messaging has been the subject of numerous digital forensic studies since the mid 2000's. In a series of early works, Dickson identified that artefacts of the client-based American Online Messenger version 5.5 (AIM) [16], MSN Messenger version 7.5 [36], Yahoo Messenger version 7.0 [37], and Trillian version 3.1 [38] could be recovered from the registry, user settings, and other application-specific files on the hard drive of a Windows XP machine. By applying keyword search, the author was able to recover portion of the conversation history from unstructured datasets such as memory dumps, slack space, free space, and swap files in plain text, even with the absence of chat logging. The findings were echoed by several others studies with respect to Digsby [39–41], Windows Live Messenger 8.0 [42], and Pidgin 2.0 [43]. However, Levendoski et al. [44] concluded that artefacts of the Yahoo Messenger client produced a different directory structure on Windows Vista/7. Kiley et al. [19] investigated web-based IM apps (i.e., AIM Express, Google Talk, Meebo, and E-Buddy) and found that artefacts of the contact lists, conversations, and approximate time of the last conversation could only be recovered from memory dump and hard disk's free space, although reference to the URLs, last access times, and view count information could be recovered from the web browsing history.

Wong et al. [45] and Al Mutawa et al. [46] demonstrated that artefacts of the Facebook web-application could be recovered from memory dumps and web browsing cache in Javascript Object Notation (JSON) and Hypertext Markup Language (HTML) formats. Al Mutawa et al. [46] also described a methodology for investigating the Arabic string artefacts on a computer device. In another study, Al Mutawa et al. [47] investigated the artefacts of the Facebook and several other IM applications on iPhone 4, Blackberry Torch 9800, and Samsung GT-i9000 Galaxy S. The authors were able to extract records of the contact list and conversation from the logical images, with the exception of the BlackBerry devices.

Said et al. [48] investigated Facebook and other IM applications for iPhone 3G and 3GS, Blackberry Bold 7000 and 900, Samsung Omnia II i8000, Nokia E71, and Ericsson G900. Of all the mobile devices investigated, it was determined that only BlackBerry Bold 9700 and iPhone 3G/3GS provided evidence of Facebooking unencrypted. The study also revealed that artefacts of the Facebook applications were unique to the mobile devices investigated (i.e., iPhone 3GS and iphone 3G had the same version of Facebook v3.4.2 but maintained different files in the application folders). Walnycky et al. [49] added that artefacts of the Facebook Messenger could vary depending on user settings, OS version, and manufacturer. Levinson et al. [50]

demonstrated that records of the recent Facebook chats stored in the property list of the Facebook Messenger for iOS can assist forensic practitioners with timeline analysis.

Examining iTunes backups rather than disk images, Norouzizadeh et al. [10] and Tso et al. [51] concluded that it is possible to extract users' personal data, messages, contact lists and posts Facebook app from the iTunes backup of iPhone 4 and iPhone 5s, respectively. Chu et al. [52] focused on live data acquisition from the desktop personal computer (PC) and was able to identify distinct strings that will assist forensic practitioners with reconstruction of the previous Facebook sessions. Wongyai and Charoenwatana [53] determined that objects recovered from a network analysis of Facebook homepage can be broadly categorised into 24 types based on properties such as file type, naming pattern, IP address, and location or section on the page.

Sgaras et al. [54] analysed Skype and several other VoIP applications for iOS and Android platforms. Although footprints of the installations, user profiles, conversations, contact lists, and network traffic could be located for all the VoIP applications investigated, it was concluded that the Android apps store far less artefacts than of the iOS apps. Simon and Slay [55] found that remnants of Skype communication, communication history, contacts, passwords, and encryption keys could be recovered from physical memory dump. However, Teng and Lin [56] demonstrated that using SQLite editor tools, one could easily modify Skype log files. Unsurprisingly, other studies have suggested that the network traffic behaviour varies among different versions [57, 58].

In the only article on Windows Store apps for instant messaging (at the time of this research), Lee and Chung [34] studied the third party Viber and Line apps and identified that the package identifications (IDs) could be discerned from '2414F_C7A.ViberFreePhoneCall-sText_p61zvh252yqyr' and 'NA_VER.LINEwin8_8ptj331gd3tyt' respectively. By analysing the app caches, the authors managed to locate records of account logins, contacts, chats, transferred file unencrypted. However, the study is only limited to dead analysis of the hard disk. Hence, there is a need to develop a further understanding of the implications of the Windows Store apps for IM forensics–a gap that this paper aims to contribute to.

## 3. Research Methodology

The examination procedure in this research is adapted from the four-stage digital forensic framework of McKemmish [59], namely: identification of digital evidence, preservation of digital evidence, analysis, and presentation. The purpose is to enable acquisition of realistic data similar to that found in real world investigations. This paper mainly focuses on the analysis stage, although we also briefly discuss the evidence source identification, preservation, and presentation to demonstrate how the framework could be applied in practice.

The first step of the experiment involved the creation of eight (8) fictional accounts to play the role of suspects and victims in this research–see Table 1. The IM accounts were assigned with a unique 'display icon' and username which was not used within the respective IM apps and Windows operating system. This eases identification of the user roles. Next was to create the test environments for the suspects and the victims, which consisted two (2) control base VMware Workstations (VMs) version 9.0.0 build 812388 running Windows 8.1 Professional (Service Pack 1, 64 bit, build 9600). As explained by Quick and Choo [23–25], using physical hardware to undertake setup, erasing, copying, and re-installing would have been an onerous exercise. Moreover, a virtual machine allows room for error by enabling the test environment to be reverted to a restore point should the results are unfavourable. The workstations were configured with the minimal space (2GB of physical memory and 20GB hard drive space) in order to reduce the time required to analyse the considerable amounts of snapshots in the latter stage.

**Table 1. Account details for IM experiments.**

| IM Experiment | Username | Email | Role |
|---|---|---|---|
| **Facebook Messenger Forensics** | John Adkins | johnadkins109@gmail.com | Suspect |
| | Jack Jeffry | fbcchelper@gmail.com | Suspect 2 |
| | Adam Jacobs | adamjacobs717@gmail.com | Victim 1 |
| | Samuel Traviss | samueltraviss@gmail.com | Victim 2 |
| | Kelvin Sky | fbcctester@gmail.com | Victim 3 |
| **Skype Forensics** | Adam Thomson/ adam.thomson84 | adamthomson1984@gmail.com | Suspect |
| | Harold Cornwall/ harold.cornwall34 | haroldcwall@gmail.com | Victim 1 |
| | Alicia Richardson/ alicia.rich19 | aliciarich19@gmail.com | Victim 2 |

doi:10.1371/journal.pone.0150300.t001

In the third step, we conducted a predefined set of activities to simulate various real world scenarios of using the apps on each workstation/test environment. The base assumptions are that the practitioner encounters a live system running Microsoft Windows 8.1 in a typical home environment. Similar to the approaches of Quick and Choo [23–25], the 3111th email message of the University of California (UC) Berkeley Enron email dataset (downloaded from http://bailando.sims.berkeley.edu/enron_email.html on 24th September 2014) was used to create the sample files and saved as SuspectToVictim.rtf, SuspectToVictim.txt, SuspectToVictim. docx, SuspectToVictim.zip, SuspectToVictim.jpg (printscreen), VictimToSuspect.rtf, Victim­ToSuspect.txt, VictimToSuspect.docx, VictimToSuspect.jpg (printscreen), and VictimToSus­pect.zip to simulate the transferring and receiving of files of different formats using the IM apps. As the filenames suggest, the 'SuspectToVictim' (and 'VictimToSuspect') files were placed on the suspect's workstation (and victims' workstations respectively) and subsequently transferred to the victims' workstations (and suspect's workstation respectively).

The experiments were predominantly undertaken in NATed (where NAT stands for Net­work Address Translation) network environment and without firewall outbound restriction to represent a typical IM situation. Wireshark was deployed on the host machine to capture the network traffic from the suspect's workstation for each scenario. After each experiment was carried out, we saved a copy of the network capture file in.PCAP format, and acquired a bit­stream (dd) image of the virtual memory (.VMEM) file prior to shutdown. We then took a snapshot of each workstation after being shutdown and made a forensic copy of the virtual disk (.VMDK) file in Encase Evidence (E01) format. This resulted in the creation of fifteen (15) snapshots (each for each environment) as highlighted in Table 2, and Figs 1 and 2. The decision to instantiate the physical memory dumps and hard disks with the virtual disk and memory files was to prevent the datasets from being modified with the use of memory/image acquisition tools [23, 25].

The final step of this research was to analyse the datasets using a range of forensically recog­nised tools (as highlighted in Table 3) and present the findings. Both indexed and non-indexed as well as Unicode and non-Unicode string searches were included as part of the evidence searches. The experiments were repeated at least thrice (at different dates) to ensure consis­tency of findings.

## 4. Analysis of the Facebook App

Facebook (Messenger) is an IM service offered by Facebook–one of the most popular social network platforms with more than one billion daily active users on average [60]. The Store app was officially released on 17th October 2013 in conjunction with the launch of Windows 8.1 [61]. It allows users to view status updates, news feeds, send and receive text and voice, as well

**Table 2. Details of VM snapshots created for this research.**

| IM forensics | Snapshot | Description |
|---|---|---|
| | **1.0 Base-Snapshot** | A control base snapshot was made to create the control media to determine changes from each IM scenario. |
| **Facebook forensics** | **F1.1 Install-Snapshot** | Using a duplicate copy of the control base snapshot (1.0), we accessed the Windows Store to download and subsequently install the Facebook app version 1.4.0.9. |
| | **F1.1.1 Login-Snapshot** | A snapshot was made of the install snapshot (F1.1) to examine the artefacts from the Facebook login. |
| | **F1.1.2 Friend-Snapshot** | A second snapshot was created of the install snapshot (F1.1) to examine the process of searching and adding friend using the Facebook app. |
| | **F1.1.3 Chat-Snapshot** | Another snapshot was made of the install snapshot (F1.1) to undertake scripted conversations and file transfers using the Facebook app. The conversations were limited to two participants. |
| | **F1.1.3.1 Uninstall-Snapshot** | A snapshot was made of the chat snapshot (F1.1.3) to examine the data remnants left behind after uninstalling the Facebook app. The app was uninstalled using the uninstall function on the start screen. |
| | **F1.1.4 Group Chat-Snapshot** | A final snapshot was made of the install snapshot (F1.1) to examine the artefacts left by the group chat feature of the Facebook app. The suspect's account was used to add all the victims into a group chat namely 'DeviGroup'. A mock conversation was subsequently taken between the suspect and the victims. |
| **Skype forensics** | **S1.1 Install-Snapshot** | Using a duplicate copy of the control base snapshot (1.0), we updated the Skype app to version 3.1.0.1007 (the latest version at the time of this research). |
| | **S1.1.1 Login-Snapshot** | A snapshot was made of the install snapshot (S1.1) to examine the login artefacts of the Skype app. |
| | **S1.1.2 Contact-Snapshot** | A second snapshot was made of the install snapshot (S1.1) to examine the process of adding contact using the Skype app. The contacts were subsequently synced to the Windows Live (login) account to identify additional artefacts in relation to the contact syncing. |
| | **S1.1.3 IM-Snapshot** | A third snapshot was made of the install snapshot (S1.1) to undertake scripted IM conversations and file transfers using the Skype app. The conversations were limited to two participants. |
| | **S1.1.3.1 Uninstall-Snapshot** | We duplicated the IM snapshot (S1.1.3) to examine the data remnants left behind after uninstalling the Skype app. Uninstallation was undertaken using the uninstall function on the start screen. |
| | **S1.1.4 Group Chat-Snapshot** | Another snapshot was created of the install snapshot (S1.1) to examine the group chat artefacts of the Skype app. The suspect's account was used to add all the victims into a group chat namely 'DeviSkypeGroup'. A mock conversation was subsequently taken between the suspect and the victims. |
| | **S1.1.5 Voice and Video Call-Snapshot** | Additional copy of the install snapshot (S1.1) was made to examine the process of voice and video calling using the Skype app. We first made a Skype to Skype voice call from the suspect to victim, followed by a video call during the voice call. |
| | **S1.1.6 Video Message-Snapshot** | A final copy of the install snapshot (S1.1) was made to investigate the process of creating video message using the Skype app. A video message was made and subsequently sent from the suspect's VM to the victim's VM. |

doi:10.1371/journal.pone.0150300.t002

as features such as file transfer and image sharing. In this section, we present artefacts of installation, uninstallation, logins, contact lists, conversations, transferred files, and notifications of the Facebook app (version 1.4.0.9) on Windows 8.1.

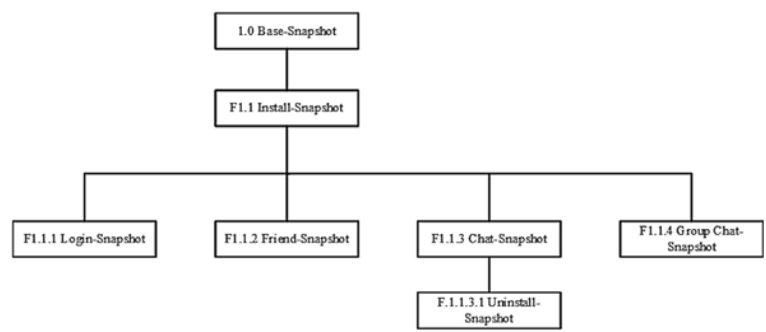

**Fig 1. VM snapshots created for Facebooking experiments.**

doi:10.1371/journal.pone.0150300.g001

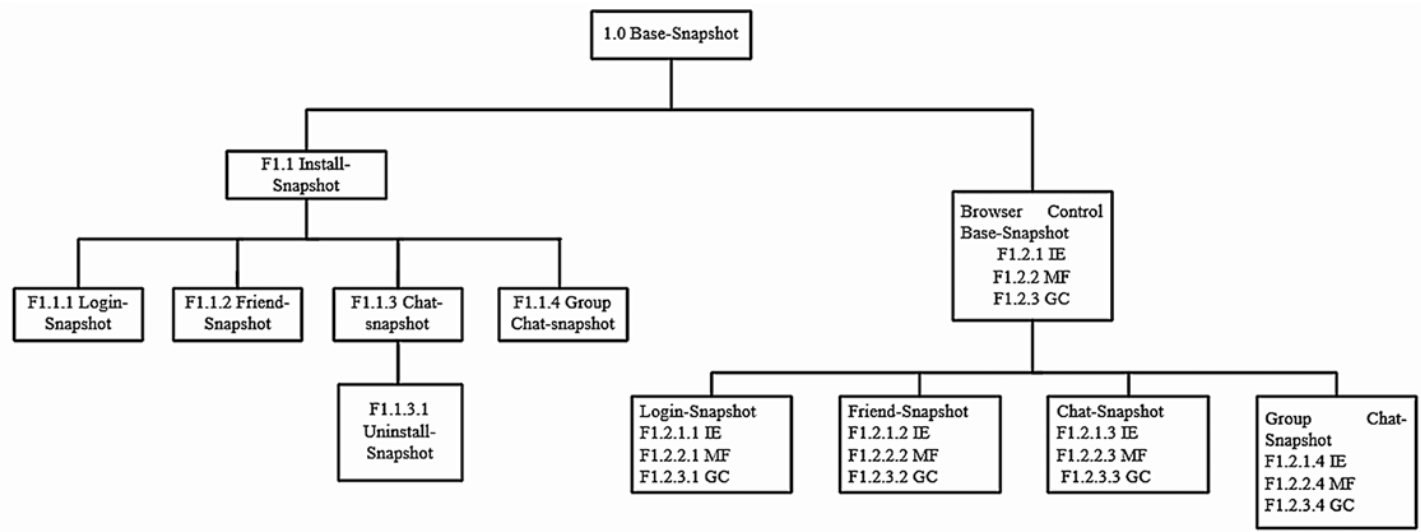

**Fig 2. VM snapshots created for Skype experiments.**

## 4.1 Installation of the Facebook App

Examinations of the directory listings observed that the package ID (for the Facebook app) can be differentiated from 'Facebook.Facebook_1.4.0.9_x64__8xx8rvfyw5nnt'. A closer examination of the registry entries created during the installation observed that the installation time could be identified from the 'InstallTime' entry within the *HKEY_USERS\<SID>\Software \Classes\Local Settings\Software\Microsoft\Windows\CurrentVersion\AppModel\Repository \Families\Faceook.Facebook_8xx8rvfyw5nnt\Facebook.Facebook_1.4.0.9_x64_8xx8rvfyw5nnt* branch in 64-bit FILETIME Hex value in Big Endian format.

A search for the package ID 'Facebook.Facebook_1.4.0.9_x64__8xx8rvfyw5nnt' in the Windows Store logs (resided at *%AppData%\Local\Temp\winstore.log* and *%AppData%\Local \Packages\winstore_cw5n1h2txyewy\AC\Temp\winstore.log*) located supporting timestamp information such as the dates when the app was first launched and updated. Moreover, analysis of the prefetch files revealed the last run time and number of times the app has been loaded in 'FACEBOOK.EXE.pf'. As for event logs, there was additional timestamp information which indicated the accessed times in 'Application.evtx', 'Microsoft-WS-Licensing%4Admin.evtx', 'Microsoft-Windows-AppModel-Runtime%4Admin.evtx', 'Microsoft-Windows-AppXDe-ploymentServer%4Operational.evtx', 'Microsoft-Windows-Audio%4PlaybackManager.evtx', 'Microsoft-Windows-CoreApplication%4Operational.evtx', 'Microsoft-Windows-PushNotifi-cation-Platform%4Operational.evtx', 'Microsoft-Windows-Resource-Exhaustion-Resolver% 4Operational.evtx', 'Microsoft-Windows-SettingSync%4Debug.evtx', 'Microsoft-Windows-Shell-Core%4Operational.evtx', 'Microsoft-Windows-TWinUI%4Operational.evtx', 'Micro-soft-Windows-Windows Firewall With Advanced Security%4Firewall.evtx', and 'System.evtx'.

Examinations of the running processes using the 'pslist' function of Volatility determined that the process name could be discerned from 'Facebook.exe'. Fig 3 illustrates that the 'pslist' output also included the process identifier (PID), parent process identifiers (PPID), and the process initiation and termination time. The PID could prove useful for correlating data associated with the the app during further analysis of the RAM (i.e., contextualising a string using the 'Yarascan' function of Volatility).

**Table 3. Tools used for IM analysis on Windows 8.1.**

| Tool | Usage |
| --- | --- |
| **FTK Imager Version 3.2.0.0** | To create forensic images for the.VMDK files. |
| **dd version 1.3.4–1** | To produce a bit-for-bit image of the.VMEM files. |
| **Autopsy 3.1.1** | To parse the file system, produce directory listings, as well as extracting or analysing stored files, browsing history, 'NTUSER.dat' registry files (using the RegRipper plugin), 'pagefile.sys' Windows swap file, and unallocated spaces located within the forensics images of VMDK files. |
| **HxD Version 1.7.7.0** | To conduct keyword searches in the unstructured datasets. |
| **Volatility 2.4** | To analyse the running processes (using the 'pslist' function), network statistics (using the 'netscan' function), and detecting the location of a string (using the 'yarascan' function) recorded in the physical memory dumps. |
| **Photorec 7.0** | To data carve the unstructured datasets. |
| **Skype Chatsync Reader** | To analyse the content of Skype's 'Chatsync' file. |
| **SQLite Browser Version 3.4.0** | To view the contents of SQLite database. |
| **Wireshark version 1.10.1** | To analyse the network traffic. |
| **Network Miner version 1.6.1** | To analyse and data carve the network files. |
| **Whois command** | To determine the registration information of the IP addresses. |
| **Nirsoft Web Browser Passview 1.19.1** | To recover the credential details stored within web browsers. |
| **Nirsoft cache viewer, ChromeCacheView 1.56, MozillaCacheView 1.62, IECacheView 1.53** | To analyse the web browsing cache. |
| **BrowsingHistoryView v.1.60** | To analyse the web browsing history. |
| **Thumbcacheviewer Version 1.0.2.7** | To examine the Windows thumbnail cache. |
| **Windows Event Viewer Version 1.0** | To view the Windows event logs. |
| **Windows File Analyser 2.6.0.0** | To analyse the Windows prefetch and link files. |
| **NTFS Log Tracker V1.2** | To parse and analyse the $LogFile, $MFT, and $UsnJrnl New Technology File System (NTFS) files. |

## 4.2 Logins

In our experiments, it was observed that Facebook maintains a wealth of cache data for the Store app in a number of SQLite databases located in *%AppData%\Local\Packages\Facebook. Facebook_1.4.0.9_x64__8xx8rvfyw5nnt\LocalState\<User specific Facebook ID>\DB\*, such as Analytics.sqlite, FriendRequests.sqlite, Friends.sqlite, Messages.sqlite, Notifications.sqlite, and Stories.sqlite. However, it is noteworthy that these databases will only appear when the user is logged in from the app. The database of interest with the logins is Analytics.sqlite, which contains records of the login time in Unix epoch format. The records can be discerned from the 'name' and 'module' table columns which reference 'login' and 'login_events' in the 'analytics_logs' table, respectively—see Fig 4. Within *%AppData%\Local\Packages\Facebook.Facebook_8xx8rvfyw5nnt\AC\InetCache\<Cache ID>\* and *%AppData%\Local\Packages \Facebook.Facebook_8xx8rvfyw5nnt\AC\.local_cache\* there were copies of profile and cover pictures of the user and the contacts, as well as other pictures which appeared on the Facebook timelines. The pictures may provide invaluable leads that lay the groundwork for follow-up via traditional investigative techniques.

| Offset(V) | Name | PID | PPID | Thds | Hnds | Sess | Wow64 | Start | Exit |
|-----------|------|-----|------|------|------|------|-------|-------|------|
| 0xfffffe0000114a900 | svchost.exe | 628 | 568 | 10 | 0 | 0 | 0 | 2014-12-13 15:29:38 UTC+0000 | |
| 0xfffffe00025757c0 | Facebook.exe | 3400 | 628 | 39 | 0 | 1 | 0 | 2015-01-19 16:28:08 UTC+0000 | |

**Fig 3. The 'pslist' output for the Facebook app.**

doi:10.1371/journal.pone.0150300.g003

A search for the login password produced no matches in the forensic image and memory dump. An examination of the network traffic revealed that the host first established a session with Symantec Certification Authority (i.e., IP address 23.58.43.27) for certificate authentication. Afterwards, the host accessed the nearest Akamai content delivery servers (i.e., IP addresses 23.62.109.*) and Facebook servers from different countries (i.e., IP addresses 31.13.*.* and 115.164.13.* in our research) on port 443 (hence HTTPS), which we theorised to retrieve the profile and timeline information. Although the network traffic was encrypted and the login credentials were not recovered, we were able to correlate the IP addresses with the timestamp information to determine when the app was started up and the duration of Facebook use in our research.

## 4.3 Friend Lists

Contact (or 'friend' in the context of Facebook) lists can be a useful reference point for a suspect's social network. A search for the suspect's profile name in the directory listing determined that artefacts of the contact lists can only be located in the Friends.sqlite database. The table of particular interest is the 'friends' table, which holds a list of user identifications (UIDs), full names, first names, middle names, last names, email addresses, phone numbers, profile links, communication rank (frequency of communication), and birth dates associated with the friends added by the user as shown in Fig 5. Moreover, the 'profiles' table provide supplementary information relating to the profiles viewed by the user such as the profile type (private profile or page), description (if any), URLs to the profiles, cover photo metadata (i.e., photo IDs, sizes, URLs, titles, and creation times for the cover photos), number of mutual friends associated with the profiles (if any), whether a friend request can be sent to the profiles, and the user has liked the page or is a subscriber.

## 4.4 Conversations and Transferred Files

Facebook allows users to transfer files up to 15MB. When a file is uploaded using the chat window, it will be attached alongside the line of chat messages (if any) and appear as a download link. The sender is allowed to abort a transfer part way through the process. The downloaded files were saved under *%Downloads%\* by default, all of which were given an Alternate Data Stream (ADS) ZoneTransfer marker (ZoneID) with reading 'ZoneID = 3', indicating that the files were downloaded from an Internet zone [62]. This also suggests that when a user downloads a file using the Facebook app, there will be records remaining in Windows system files such as $LogFile, $MFT, and $UsnJrnl to indicate the filenames, directory paths, and timestamps for the downloaded files; an excerpt of the $LogFile entries (recovered from the suspect's workstation) is shown in Fig 6. Analysis of the thumbnail caches stored within %

| | id | time | log_type | name | module | extra |
|---|-----|------|----------|------|--------|-------|
| | Filter | Filter | Filter | Filter | Filter | Filter |
| 1 | 1 | 1421898314666 | client_event | login | login_events | {} |

**Fig 4. Login records located in the 'analytics_logs' table of Analytics.sqlite database.**

doi:10.1371/journal.pone.0150300.g004

| uid | name | first_name | middle_name | last_name | contact_email | phones | profile_url | is_pushable | has_messenger | :ommunication_rank | birthday_date |
|---|---|---|---|---|---|---|---|---|---|---|---|
| Filter | Filter | Filter | Filter | Filter | Filter | Filter | Filter | Filter | Filter | Filter | Filter |
| 1 | 100004911219827 | Kelvin Sky | Kelvin | | Sky | fbcctester@gmail.com | [] | https://www.facebook.com/kelvin.sky.52 | 0 | 0 | 0.000840054885... | 1990-01-01 00:00:00 |

**Fig 5. The 'friends' table of Friends.sqlite database.**

doi:10.1371/journal.pone.0150300.g005

*AppData%\Local\Packages\Package ID\AC\INetCache\<Cache ID>\* and *%AppData%\Local \Microsoft\Windows\Explorer\* (henceforth thumbcache) determined that copies of the transferred or downloaded can be recovered. This creates potential for alternative methods for recovery of the deleted files, but the results may not be definitive.

Examinations of the cache databases determined that artefacts of the conversations could be recovered from the Analytics.sqlite and Messages.sqlite databases. Within the 'analytics_logs' table of the former there were timestamp records which reflected the times when the chat tab was turned on, conversations were initiated by the user, as well as files were downloaded. The entry of which could be discerned from the 'name' table column which referenced 'chat_turned_on', 'message_sent_attempt' or 'message_send_state', and 'file_downloaded' respectively. Meanwhile, details about the conversations and file transfers were recovered from the 'messages' table in the latter. Each thread created an entry which comprised the thread ID, conversation texts (if any), UID and username of the sender and the receiver, a count of the number of times the message was sent, file attachment metadata (i.e., sender's username and ID as well as filename, file size, and format references for the files transferred as shown in Fig 7), and other relevant information as shown in Fig 8. Additionally, the 'users' table (of the Messages. sqlite database) could provide additional information pertaining to the correspondents including the UIDs, email addresses, Facebook names, last active times and other information as detailed in Fig 9.

Undertaking data carving of the memory captures and unallocated space only produced matches to the transferred/downloaded sample files. By searching for terms unique to the app cache databases (i.e., table column names), it was possible to recover complete/partial fragments of the databases in plain text (similar to other IM scenarios). However, there was no common footer information to indicate the file structure. Fig 10 illustrates that records of conversations from the 'messages' table (of Messsages.sqlite database) can be located using the table column name 'm_mid'. Moreover, we were also able to locate copies of Asynchronous JavaScript and XML (AJAX) objects for the Facebook chat in the memory captures. The

**Fig 6. $LogFile entries for the Facebook app's file download.**

doi:10.1371/journal.pone.0150300.g006

```
[{
    "name": "10934517_391924760981228_2133990913_n.jpg",
    "size": 0,
    "id": "391924760981228",
    "localurl": null,
    "height": 960,
    "width": 742,
    "preview":                                                                              "https://fbcdn-sphotos-h-a.akamaihd.net/hphotos-ak-xpa1/v/t34.0-
12/p206x206/10934517_391924760981228_2133990913_n.jpg?oh=8d69df2a97d71f567739c0e2e20e61eb&oe=54DECA5D&__gda__=1423880331_d81326e18103fd67386d5f7f38bf83f5",
    "url":                                                                                   "https://fbcdn-sphotos-h-a.akamaihd.net/hphotos-ak-xpa1/v/t34.0-
12/10934517_391924760981228_2133990913_n.jpg?oh=16498c19d5072ead455d682297f4e9fd&oe=54DEA7C8&__gda__=1423888552_385688eacbf4d81b12eedfbd13add909",
    "mime": "image/jpeg",
    "type": 4
}, {
    "name": "VictimToSuspect.pdf",
    "size": 31747,
    "id": "391924720981232",
    "localurl": null,
    "mime": "application/pdf",
    "type": 7
...
}]
```

**Fig 7. File attachment metadata recorded in the 'attachments' field of the 'messages' table.**

artefacts could provide a clear indication of contact in Unix epoch format, Facebook usernames and UIDs of the correspondents, and conversation texts as depicted in Fig 11. The JSON coding could be a suitable search keyword for future searches. The presence of the remnants in the memory space of 'Facebook.exe' confirmed that the texts were associated with the Facebook app.

Inspecting the network traffic, it was observed that the transferred files were uploaded to IP addresses 31.13.70.*, 31.13.67.*, and 31.13.67.* with URLs referencing 'upload.facebook.com'. The downloaded files were seen from IP addresses 31.13.70.*, and the URLs were prefixed with 'cdn.fbsdx.com'. Meanwhile, the IP addresses i.e., 31.13.79.* and 31.13.76.102 were observed in relation to the conversations, with URLs referencing '5-edge-chat.facebook.com'—see Table 4 for details. Although the contents were encrypted completely, the IP addresses and URLs highlighted as part of our research may assist a practitioner in scoping the Facebook activities undertaken by a suspect in future investigations. Additionally, the IP addresses can be correlated with the 'netscan' output (of Volatility) to obtain information regarding the running process (i.e., PID, process creation time, and socket states) as detailed in Fig 12.

## 4.5 Real-time Notifications

Facebook notifications prompt users in real-time when activities such as messages and comments were posted on their walls, or wall post tagging took place. Analyses of the directory listings only revealed records of the notifications in the 'notifications' table of Notifications.sqlite database. The records contained the senders' UIDs, notification texts, URLs, update and creation times, whether a notification has been read by the user ('1' for read and '0' for unread), and other options useful to aid timeline analysis (see Fig 13).

| | id | thread_id | body | sender | tags | timestamp | tion_ | line_ | rdin | attachments |
|---|---|---|---|---|---|---|---|---|---|---|
| | Filter | Filter | Filter | Filter | Filter | Filter | Fil... | Fil... | Fil... | Filter |
| 15 | m_mid.1421... | t_msg.e241a... | Here are some files for you ... | {"user_id":"100004911219827","name":"Kelvin Sky","e... | ["inbox","read","sent","source:chat"] | 1421644786450 | 14... | N... | N... | [] |
| 16 | m_mid.1421... | t_msg.e241a... | | {"user_id":"100004935817781","name":"Jack Jeffry","e... | ["inbox","read","source:chat"] | 1421644776796 | 14... | N... | N... | [{"name":"10934350_38423279175124... |
| 17 | m_mid.1421... | t_msg.e241a... | Here are some files for you | {"user_id":"100004935817781","name":"Jack Jeffry","e... | ["inbox","read","source:chat"] | 1421644752737 | 14... | N... | N... | [] |
| 18 | m_mid.1421... | t_msg.e241a... | Hello Victim | {"user_id":"100004935817781","name":"Jack Jeffry","e... | ["inbox","read","source:chat"] | 1421644744425 | 14... | N... | N... | [] |
| 19 | m_mid.1421... | t_msg.e241a... | Hello Suspect | {"user_id":"100004911219827","name":"Kelvin Sky","e... | ["inbox","read","sent","source:chat"] | 1421644736327 | 14... | N... | N... | [] |

**Fig 8. The 'messages' table of Messages.sqlite database.**

| | id | type | email | name | first_name | middle_name | last_name | pushable | has_messenger | st_active_timestan | last_active_active | last_active_source |
|---|---|---|---|---|---|---|---|---|---|---|---|---|
| | Filter | Filter | Filter | Filter | Filter | Filter | Filter | Filter | Filter | Filter | Filter | Filter |
| 1 | 100004911219827 | NULL | 1000049112198... | Kelvin Sky | Kelvin | | Sky | 1 | 1 | 1423766054 | 1 | |
| 2 | 100004935817781 | NULL | 1000049358177... | Jack Jeffry | Jack | | Jeffry | 0 | 0 | 1423766043 | 1 | |

**Fig 9. The 'users' table of Messages.sqlite database.**

## 4.6 Uninstallation of the Facebook App

Uninstallation of the Facebook app did not create uninstallation files. When the uninstallation was taken place, only the installation folder remained, but was moved to *%Program Files% \WindowsApps\Deleted*. Other footprints such as remnants from RAM, unallocated space, and system files such as pagefile.sys, shortcuts, event logs, prefetch files, $LogFile, $MFT, as well as $UsnJrnl were not affected by uninstallation process. The uninstallation also created additional references to the directory paths and timestamp information for the files removed during the uninstallation in $LogFile, $MFT, as well as $UsnJrnl.

## 5. Analysis of the Skype App

Skype is a popular IM and Voice over Internet Protocol (VoIP) application that provides free IM services, audio and video calls between computers and other mobile devices [63]. With the recent launch of Windows 8.1, Skype is now an integrated Windows service. The most recent version of Skype uses the Super Wideband Audio Codec (SILK) [64]. The overlay peer-to-peer network consists of a combination of ordinary and supernodes [57]. An ordinary node is a typical Skype application that provides the users the ability to place calls and send text messages. The supernode serves as a proxy to relay information between nodes with firewall restrictions and an intermediary to handle authentication and user lookups during logins [57].

In this section, we present results of our investigation of artefacts left behind after the use of the Skype (Windows store) app version 3.1.0.1007 on Windows 8.1, such as installation directory paths, usernames, passwords, text of conversations, transferred or downloaded files, records of video and voice calls, and the associated timestamps.

## 5.1 Installation of the Skype App

Analysis of the directory listing identified that the package ID could be discerned from 'Microsoft.SkypeApp_kzf8qxf38zg5c'. The package ID was then used to correlate the 'InstallTime' registry entry, Windows Store logs, and event logs to determine the installation and accessed

```
Owner: Process Facebook.exe Pid 3400
0x37f869d488  6d 5f 6d 69 64 2e 31 34 32 31 36 34 33 38 30 30   m_mid.1421643800
0x37f869d498  31 39 32 3a 35 62 37 35 62 66 31 62 66 31 38 31   192:5b75b1bf181
0x37f869d4a8  65 35 37 34 32 74 5f 6d 73 67 2e 65 32 34 31   e57424t_msg.e241
0x37f869d4b8  61 66 35 38 61 39 37 65 31 30 36 31 61 35 37 64   af58a97e1061a57d
0x37f869d4c8  66 39 35 62 32 64 30 34 38 30 64 33 32 48 65   f95b2d0480d352He
0x37f869d4d8  72 65 2e 61 72 65 2e 73 6f 6d 65 2e 66 69 6c 65   re.are.some.file
0x37f869d4e8  73 20 66 6f 72 20 79 6f 75 20 53 55 53 50 45 43   s.for.you.SUSPEC
0x37f869d4f8  54 7b 22 75 73 65 72 5f 69 64 22 3a 22 31 30 30   T{"user_id":"100
0x37f869d508  30 30 34 39 31 31 32 31 39 38 32 37 22 2c 22 6e   004911219827","n
0x37f869d518  61 6d 65 22 3a 22 4b 65 6c 76 69 6e 20 53 6b 79   ame":"Kelvin.Sky
0x37f869d528  22 2c 22 65 6d 61 69 6c 22 3a 22 31 30 30 30 30   ","email":"10000
0x37f869d538  34 39 31 31 32 31 39 38 32 37 40 66 61 63 65 62   4911219827@faceb
0x37f869d548  6f 6f 6b 2e 63 6f 6d 22 7d 5b 22 69 6e 62 6f 78   ook.com"}["inbox
0x37f869d558  22 2c 22 72 65 61 64 22 2c 22 73 6f 75 72 63 65   ","read","source
0x37f869d568  3a 63 68 61 74 22 5d 2e 4b 00 92 de 91 13 ba b1   :chat"].K......
0x37f869d578  81 0d 21 5a 80 5b 5d 01 4b 00 92 de 91 8b 35 0f   ..!Z.[].K....5.
```

**Fig 10. Portion of the 'messages' table of Messages.sqlite database recovered from the memory space of 'Facebook.exe'.**

Time: 2015-01-19T16:36:24Z...Channel: 1;12401170449359906421...Type: wns/raw..Msg-Id: 4A0DD6A11DD910AE..Ack: true

```
{
    "message": "Kelvin Sky: Here are some files for you SUSPECT",
    "time": 1421685383,
    "is_logged_out_push": false,
    "target_uid": 100004935817781,
    "params": {
        "action_id": "1421685383451000000",
        "uid": "100004911219827",
        "push_phase": "V3",
        "disable_light": "1",
        "PushNotifID": "77bf53eb-8fc8-4b63-b3f9-13eb240cd7b3",
        "disable_vibrate": "1",
        "disable_sound": "1",
        "tid": "4397584927746659",
        "d": "b272fb5G5af436acda35G0G0G91da71d",
        "a": "100004911219827",
        "n": "mid.1421685383423:002f4592b91a2c3779",
        "o": "1",
        "m": "2ef9adeb98",
        "u": "100004935817781",
        "t": "2541960aad",
        "s": "1421685383437",
        "unified_tid": "msg.e241af58a97e1061a57df95b2d0480d352"
    },
    "type": "orca_message",
    "unread_count": 1
}
```

**Fig 11. Remnants of Facebook chat recovered from suspect's RAM in JSON.**

doi:10.1371/journal.pone.0150300.g011

times. An inspection of the prefetch files determined that the process name (for the Skype app) was masqueraded with 'WWAHost.exe'—the process name for the Store apps written in JavaScript [35]. As the same process name was located for more than one app of the same type, it was not possible to determine exactly which prefetch file was associated with the Skype app.

**Table 4. Network information observed for the Facebook app.**

| Registered owner | IP address(es) | URL(s) observed |
|---|---|---|
| Akamai Technologies Inc. | 23.58.43.27 | e8218.ce.akamaiedge.net, ocsp.ws.symantec.com.edgekey.net, gtssl-ocsp.geotrust.com, g.symcd.com |
| Akamai Technologies Inc. | 23.62.109.216 | a2047.dspl.akamai.net, fbcdn-profile-a.akamaihd.net |
| Akamai Technologies Inc. | 23.62.109.87 | a591.dspda2.akamai.net, fbcdn-vthumb-a.akamaihd.net.edgesuite.net |
| Facebook Malaysia | 31.13.67.7, 31.13.67.23 | scontent-a-kul.xx.fbcdn.net |
| Facebook USA | 31.13.70.1 | star.c10r.facebook.com, api.facebook.com, www.facebook.com, star.facebook.com, upload.facebook.com |
| Facebook Singapore | 31.13.79.246 | star.c10r.facebook.com, api.facebook.com, star.facebook.com, 5-edge-chat.facebook.com, upload.facebook.com, www.facebook.com |
| Facebook USA | 31.13.70.7 | scontent.xx.fbcdn.net, cdn.fbsbx.com |
| Facebook USA | 31.13.76.102 | star.c10r.facebook.com, 5-edge-chat.facebook.com |
| DiGi Telecommunications Sdn Bhd | 115.164.13.20 | a1854.dspmm1.akamai.net, fbcdn-photos-e-a.akamaihd.net.edgesuite.net, a1073.dsw4.akamai.net, fbcdn-creative-a.akamaihd.net.edgesuite.net |
| DiGi Telecommunications Sdn Bhd | 115.164.13.25 | a1168.dsw4.akamai.net, fbstatic-a.akamaihd.net.edgesuite.net, a1531.dsw4.akamai.net, fbexternal-a.akamaihd.net.edgesuite.net, a1170.dsw4.akamai.net, fbcdn-dragon-a.akamaihd.net.edgesuite.net, a1854.dspmm1.akamai.net, fbcdn-photos-e-a.akamaihd.net.edgesuite.net |
| DiGi Telecommunications Sdn Bhd | 115.164.141.10 | a1005.dspw42.akamai.net, fbcdn-sphotos-a-a.akamaihd.net.edgesuite.net, a1005.dspw42.akamai.net, fbcdn-sphotos-a-a.akamaihd.net.edgesuite.net |
| DiGi Telecommunications Sdn Bhd | 115.164.141.16, 115.164.141.17 | a1406.dspw42.akamai.net, fbcdn-sphotos-f-a.akamaihd.net.edgesuite.net, a1406.dspw42.akamai.net, fbcdn-sphotos-f-a.akamaihd.net.edgesuite.net |
| DiGi Telecommunications Sdn Bhd | 115.164.141.32, 115.164.141.34, 115.164.141.40 | a1003.dspw41.akamai.net, fbcdn-sphotos-c-a.akamaihd.net.edgesuite.net, a1404.dspw41.akamai.net, fbcdn-sphotos-d-a.akamaihd.net.edgesuite.net, a1408.dspw43.akamai.net, fbcdn-sphotos-h-a.akamaihd.net.edgesuite.net |
| Facebook Inc. | 173.252.103.16 | orcart.vvv.facebook.com, orcart.facebook.com |
| Facebook Inc. | 173.252.120.6 | www.facebook.com |

doi:10.1371/journal.pone.0150300.t004

| Offset(P) | Proto | Local Address | Foreign Address | State | Pid | Owner | Created |
|---|---|---|---|---|---|---|---|
| 0x7b427010 | UDPv4 | 0.0.0.0: | *:* | | 3400 | Facebook.exe | 2015-01-19 16:29:21 UTC+0000 |
| 0x7b77bec0 | UDPv4 | 0.0.0.0: | *:* | | 3400 | Facebook.exe | 2015-01-19 16:29:21 UTC+0000 |
| 0x7b77bec0 | UDPv6 | 0.0.0.0: | *:* | | 3400 | Facebook.exe | 2015-01-19 16:29:21 UTC+0000 |
| 0x7b295d10 | TCPv4 | 192.168.220.176:49402 | 31.13.70.1:443 | ESTABLISHED | 3400 | Facebook.exe | |
| 0x7b5273b0 | TCPv4 | 192.168.220.176:49431 | 31.13.73.246:443 | ESTABLISHED | 3400 | Facebook.exe | |
| 0x7b52b360 | TCPv4 | 192.168.220.176:49364 | 23.62.109.185:443 | CLOSED | 3400 | Facebook.exe | |
| 00x7b966430 | TCPv4 | 192.168.220.176:49423 | 31.13.70.1:443 | ESTABLISHED | 3400 | Facebook.exe | |
| 0x7b972190 | TCPv4 | 192.168.220.176:49396 | 23.62.109.160:443 | ESTABLISHED | 3400 | Facebook.exe | |
| 0x7bd70390 | TCPv4 | 192.168.220.176:49394 | 31.13.79.246:443 | CLOSE_WAIT | 3400 | Facebook.exe | |
| 0x7c06f460 | TCPv4 | 192.168.220.176:49405 | 31.13.70.1:443 | ESTABLISHED | 3400 | Facebook.exe | |
| 0x7c3a3600 | TCPv4 | 192.168.220.176:49426 | 115.164.141.40:443 | CLOSED | 3400 | Facebook.exe | |
| 0x7d79d920 | UDPv4 | 0.0.0.0: | *:* | | 3400 | Facebook.exe | 2015-01-19 16:29:21 UTC+0000 |
| 0x7d70bd10 | TCPv4 | 192.168.220.176:49395 | 31.13.70.1:443 | ESTABLISHED | 3400 | Facebook.exe | |
| 0x7d77c010 | TCPv4 | 192.168.220.176:49392 | 23.62.109.160:443 | ESTABLISHED | 3400 | Facebook.exe | |
| 0x7d7d6010 | TCPv4 | 192.168.220.176:49432 | 115.164.141.32:443 | ESTABLISHED | 3400 | Facebook.exe | |
| 0x7d80a580 | TCPv4 | 192.168.220.176:49391 | 31.13.79.246:443 | ESTABLISHED | 3400 | Facebook.exe | |
| 0x7d91fd10 | TCPv4 | 192.168.220.176:49424 | 31.13.70.1:443 | CLOSE_WAIT | 3400 | Facebook.exe | |
| 0x7d9cb420 | TCPv4 | 192.168.220.176:49430 | 31.13.70.1:443 | ESTABLISHED | 3400 | Facebook.exe | |
| 0x7dc0cd70 | UDPv4 | 0.0.0.0: | *:* | | 3400 | Facebook.exe | 2015-01-19 16:29:21 UTC+0000 |
| 0x7dc0cd70 | UDPv6 | 0.0.0.0: | *:* | | 3400 | Facebook.exe | 2015-01-19 16:29:21 UTC+0000 |
| 0x7dd2ed10 | TCPv4 | 192.168.220.176:49408 | 31.13.79.246:443 | CLOSED | 3400 | Facebook.exe | |
| 0x7dd59260 | TCPv4 | 192.168.220.176:49425 | 115.164.141.40:443 | CLOSED | 3400 | Facebook.exe | |

**Fig 12. The 'netscan' output for the Facebook app.**

doi:10.1371/journal.pone.0150300.g012

## 5.2 Logins

The crucial artefacts were predominantly located in the user-specific *%AppData%\Local\Packages\Microsoft.SkypeApp_kzf8qxf38zg5c\LocalState\<Skype name>\main.db* database (unless otherwise stated, all tables will henceforth be referred to this database). Of particular interest with respect to the logins is the 'Accounts' table, which maintains a list of details about the Skype accounts logged in from the computer under investigation. The details comprise the account registration times in Unix epoch format, Microsoft Live usernames, Skype names, users' full name, birth dates, gender, registered locations, phone numbers, email addresses, homepage URLs (if any), mood texts and the creation times, time zones, and other information useful for user profiling. To recover the avatars used by the users, the practitioner can access *%AppData%\Local\Packages\Microsoft.SkypeApp_kzf8qxf38zg5c\LocalState\avatars\*.

Analysis of the Internet Explorer's web browsing history was able to identify two URLs associated with the logins, which were 'login.skype.com/login?message=signin_continue&return_url=. . .' and 'login.skype.com/login/sso?nonce=. . .'). The web browsing history can provide an estimate of the number of times a suspect had accessed Skype as well as the corresponding login times on the computer under investigation.

Examination of the *%AppData%\Local\Packages\Microsoft.SkypeApp_kzf8qxf38zg5c\LocalState\shared.xml* file indicated the Skype name and node ID of the user in the 'Default' and 'NodeID' tags, respectively. The Skype name can prove useful for correlating events initiated by the user during further analysis. Meanwhile, it was observed that the 'HostCache' tag maintains a string of the supernode IP addresses and port pairs that Skype builds and refreshes regularly [57]. Each of which is recorded in twelve character hexadecimal strings and prefixed with '0400050041050200' [65]. The shared.xml file also held records of the last used external IP address, port number, and last connected supernode IP address and port pair in the 'LastIP', 'ListeningPort', 'Supernode' tags in decimal format, respectively—see Fig 14; useful to support network analysis.

| | notification_id | object_id | object_type | sender_id | title_text | title_html | icon_url | href | unread | updated_time | created_time | join_data |
|---|---|---|---|---|---|---|---|---|---|---|---|---|
| | Filter | Filter | Filter | Filter | Filter | Filter | Filter | Filter | Filter | Filter | Filter | Filter |
| 1 | 19223616 | 10000493581781 | friend | 100004835817781 | Jack Jeffry acce... | <a href="http://... | https://fbstatic... | http://www.fac... | 1 | 2015-02-12 17:5... | 2015-02-12 17:4... | {"logging_data":"{\"alert... |
| 2 | 19224843 | 10000491121198... | stream | 100004835817781 | Jack Jeffry post... | <a href="http://... | https://fbstatic... | http://www.fac... | 0 | 2015-02-12 18:3... | 2015-02-12 18:3... | {"logging_data":"{\"alert... |

**Fig 13. The 'notifications' table of Notifications.sqlite database.**

doi:10.1371/journal.pone.0150300.g013

<HostCache>41C8010500410502006FDD4D9E9C560001040002B1B8F4A5050003B1B8F4A50500**0400050041050200**4137DF188109**0001040002B6D499A
<LastIP>1940151468</LastIP>
<LastProbingFailed>0</LastProbingFailed>
<ListeningPort>37439</ListeningPort>
<NatTracker>
    <ContraProbeResults/>
    <PreviousNatType>6</PreviousNatType>
</NatTracker>
<NetDetectOK>1</NetDetectOK>
<ReconnectSecret>8708501885DCBA67</ReconnectSecret>
<Supernode>**111.221.77.148:40028**</Supernode>

**Fig 14. Network information observed in shared.xml.**

Although the process name was masqueraded with 'WWAHost.exe', we could correlate the supernode IP addresses (obtained from the shared.xml file) with the 'netscan' output (of Volatility) to determine the PID. For example, when we mapped the supernode IP address of '111.221.77.148' with the 'netscan' output recovered from our research (see Fig 15), we obtained the PID '656'. The PID could then be used to map the 'pslist' output (of Volatility) to obtain additional information such as the PPID and process creation time as shown in Fig 16. Further analysis of the unstructured datasets identified that the config.xml and shared.xml files can be potentially carved from the memory dump and unallocated space using the header and footer values of "3C 3F 78 6D 6C 20 76 65 72 73 69 6F 6E 3D 22. . . 3C 2F 55 49 3E 0D 0A 3C 2F 63 6F 6E 66 69 67 3E 0D 0A" and "3C 3F 78 6D 6C 20 76 65 72 73 69 6F 6E 3D 22. . .3C 2F 4C 69 62 3E 0D 0A 3C 2F 63 6F 6E 66 69 67 3E 0D 0A" respectively, but the findings may be subject to software updates.

Upon launching the app, it was observed that the host first established a session with Edge-Cast Networks to download Microsoft's certificate revocation list (CRL) on port 80. The next session was established with the Akamai servers to retrieve the contact (i.e., IP address 23.58.236.138) and advertisement information (i.e., IP address 23.58.154.154) on port 443. Then, a session was established with the Microsoft servers (i.e., IP addresses 168.63.212.78 and 137.116.32.77 on port 443) for the traffic management service. When the logins occurred, the host first established several TCP sessions with random supernodes, which we hypothesised for user lookups [57]. Similar to the observation of Azab et al. [57], the IP addresses were associated with a combination of random and destined (33033) port numbers. The next servers accessed were the Windows Live Messenger server (i.e., IP address 65.54.184.60), Windows

| Offset(P) | Proto | Local Address | Foreign Address | State | Pid | Owner | Created |
|---|---|---|---|---|---|---|---|
| 0x23ef5990 | TCPv4 | 192.168.220.176:49806 | 175.210.53.47:14694 | ESTABLISHED | 656 | WWAHost.exe | |
| 0x2c5388c0 | TCPv4 | 192.168.220.176:49777 | 65.55.157.151:443 | CLOSED | 656 | WWAHost.exe | |
| 0x2f7e9d10 | TCPv4 | 192.168.220.176:49819 | 202.138.205.33:63321 | ESTABLISHED | 656 | WWAHost.exe | |
| 0x401d5d10 | TCPv4 | 192.168.220.176:49753 | 111.221.77.148:443 | ESTABLISHED | 656 | WWAHost.exe | |
| 0x69cd7860 | TCPv4 | 0.0.0.0:443 | 0.0.0.0:0 | LISTENING | 656 | WWAHost.exe | |
| 0x69cd7860 | TCPv6 | :::443 | :::0 | LISTENING | 656 | WWAHost.exe | |
| 0x6c36f6170 | TCPv4 | 192.168.220.176:49805 | 202.185.72.237:22045 | ESTABLISHED | 656 | WWAHost.exe | |
| 0x7084d5c0 | TCPv4 | 192.168.220.176:49820 | 210.195.27.167:2374 | CLOSED | 656 | WWAHost.exe | |
| 0x7a6056a0 | TCPv4 | 192.168.220.176:49763 | 91.190.218.62:12350 | ESTABLISHED | 656 | WWAHost.exe | |
| 0x7ac6d820 | TCPv4 | 192.168.220.176:49823 | 210.195.27.167:2374 | ESTABLISHED | 656 | WWAHost.exe | |
| 0x7c0l4cc0 | TCPv4 | 192.168.220.176:49789 | 126.4.219.9:60159 | CLOSED | 656 | WWAHost.exe | |
| 0x7cfb51b0 | TCPv4 | 192.168.220.176:49750 | 137.116.32.77:443 | CLOSED | 656 | WWAHost.exe | |
| 0x7d0fc340 | TCPv4 | 0.0.0.0:80 | 0.0.0.0:0 | LISTENING | 656 | WWAHost.exe | |
| 0x7d0fc340 | TCPv6 | :::80 | :::0 | LISTENING | 656 | WWAHost.exe | |
| 0x7d0b9a50 | UDPv4 | 0.0.0.0:* | *:* | | 656 | WWAHost.exe | 2015-01-19 15:00:54 UTC+0000 |
| 0x7d0b9a50 | UDPv6 | :::0 | *:* | | 656 | WWAHost.exe | 2015-01-19 15:00:54 UTC+0000 |
| 0x7dcf1710 | UDPv4 | 0.0.0.0:0 | *:* | | 656 | WWAHost.exe | 2015-01-19 15:00:54 UTC+0000 |
| 0x7dcf1710 | UDPv6 | :::0 | *:* | | 656 | WWAHost.exe | 2015-01-19 15:00:54 UTC+0000 |
| 0x7dd76ac0 | TCPv4 | 0.0.0.0:37439 | 0.0.0.0:0 | LISTENING | 656 | WWAHost.exe | |
| 0x7dd76ac0 | TCPv6 | :::37439 | :::0 | LISTENING | 656 | WWAHost.exe | |
| 0x7da12340 | TCPv4 | 192.168.220.176:49810 | 126.4.219.9:60159 | CLOSED | 656 | WWAHost.exe | |
| 0x7dac0910 | TCPv4 | 192.168.220.176:49816 | 191.236.104.206:443 | CLOSED | 656 | WWAHost.exe | |
| 0x7db88070 | TCPv4 | 192.168.220.176:49818 | 115.132.11.49:9240 | ESTABLISHED | 656 | WWAHost.exe | |
| 0x7ddf7620 | TCPv4 | 192.168.220.176:49803 | 118.100.246.15:35145 | ESTABLISHED | 656 | WWAHost.exe | |

**Fig 15. The 'netscan' output for the Skype app.**

| Offset(V) | Name | PID | PPID | Thds | Hnds | Sess | Wow64 | Start | Exit |
|---|---|---|---|---|---|---|---|---|---|
| 0xfffffe00000126900 | svchost.exe | 632 | 572 | 12 | 0 | 0 | 0 | 2015-01-19 13:44:06 UTC+0000 | |
| 0xfffffe000081a3c0 | WWAHost.exe | 656 | 632 | 53 | 0 | 1 | 1 | 2015-01-19 15:00:52 UTC+0000 | |

**Fig 16. The 'pslist' output for the Skype app.**

doi:10.1371/journal.pone.0150300.g016

Live servers (i.e., IP addresses 65.55.246.\*), as well as Hotmail server (i.e., IP address 65.55.68.104) on port 443 for login authentication and buddy list retrieval. The sessions were subsequently seen with random IP addresses on random UDP ports. Also observed were many connections to the IP addresses 91.190.216.\* (referencing 'rstwh.skype-cr.akadns.net' and '1007.0.1.3.9.r.skype.net') on random TCP port numbers, but we were unable to identify the actual functions of the IP addresses due to lack of information from the URLs as well as encrypted traffic—see Table 5 for details of the captured network traffic. Rebuilding the network files using Netminer, we only recovered certificates that were used to authenticate the HTTPS sites as well as HTML documents and image files from the HTTP sites. Since the network traffic was encrypted (HTTPS), no credential information was recovered from the network captures.

## 5.3 Contacts

Artefacts of the contacts were located in the 'Contacts' table. The artefacts comprised the Skype names, full names, birth dates, gender details, languages, registered locations, contact numbers, email addresses, homepage URLs (if any), mood texts, time zones, last online times, display names, last accessed times, and other information as depicted in Fig 17. Examination of the *%AppData%\Local\Packages\Microsoft.SkypeApp_kzf8qxf38zg5c\LocalState\<Skype name>\config.xml* file revealed the user ID for the contact with whom the user last communicated as well as the last accessed time. Each contact formed an opening and closing subtag in the 'u' tag as shown in Fig 18.

**Table 5. Network information observed for the Skype app.**

| Registered owner | IP address(es) | URL(s) observed |
|---|---|---|
| Akamai Technologies, Inc. | 23.58.43.27 | e8218.ce.akamaiedge.net, ocsp.ws.symantec.com.edgekey.net, ocsp.verisign.com |
| Akamai Technologies, Inc. | 23.58.236.138 | e4593.g.akamaiedge.net, wildcard.skype.com.edgekey.net |
| Akamai Technologies, Inc. | 23.58.154.154 | e8011.g.akamaiedge.net, wildcard.msads.net.edgekey.net |
| Microsoft Corp. | 65.54.184.60 | baymsgr1010611.gateway.messenger.live.com |
| Microsoft Corp. | 65.55.68.104 | activesync.glbdns2.microsoft.com, m.hotmail.com |
| Microsoft Corp. | 65.55.246.85, 65.55.246.149 | proxy-blu-people.directory.live.com.akadns.net, proxy-blu-people.directory.live.com |
| Privately Owned Enterprise "M.O.D.A." | 91.190.216.51, 91.190.216.56, 91.190.216.57, 91.190.216.58, 91.190.216.59, 91.190.216.62, 91.190.216.63, 91.190.216.66 91.90.218.52, 91.90.218.53, 91.90.218.54, 91.90.218.55, 91.90.218.56, 91.90.218.58, 91.90.218.59, 91.90.218.66 | rstwh.skype-cr.akadns.net, 1007.0.1.3.9.rst15.r.skype.net |
| CloudFlare, Inc. | 108.162.232.204, 108.162.232.199 | ocsp.globalsign.com, ocsp2.globalsign.com |
| Microsoft Corp. | 168.63.212.78, 137.116.32.77 | skypeecs-prod-ase-0.cloudapp.net, a.config.skype.trafficmanager.net |
| EdgeCast Networks, Inc. | 192.229.145.200 | cs1.wpc.v0cdn.net, az361816.vo.msecnd.net, certrevoc.vo.msecnd.net, mscrl.microsoft.com |

doi:10.1371/journal.pone.0150300.t005

| id | is_permanent | type | skypename | pstnnumber | aliases | fullname | birthday | gender | languages | country | province | city | phone_home | phone_office | phone_mobile | emails | hashed_emails | ho |
|---|---|---|---|---|---|---|---|---|---|---|---|---|---|---|---|---|---|---|
| Fil... | Filter | Filter | Filter | Filter | Filter | Filter | Filter | Filter | Filter | Filter | Filter | Filter | Filter | Filter | Filter | Filter | Filter | Filter |
| 1 20 | 1 | 1 | echo123 | NULL | NULL | Echo / Sound T... | NULL | NULL | en | NULL | NULL | NULL | NULL | NULL | NULL | NULL | ef36035bab630... | http:/ |
| 2 25 | 1 | 1 | harold.cornwall1 | NULL | NULL | Harold Cornwall | 19800202 | 1 | en | my | NULL | Malacca | NULL | NULL | +600156688796 | NULL | 0a44e8ecbf43b... | NULL |

**Fig 17. An excerpt of the 'Contacts' table of main.db database.**

doi:10.1371/journal.pone.0150300.g017

When the Skype account was synced with the Microsoft account, additional profile information was recovered for the contacts in the address book located at *%Appdata%\Local\Packages\microsoft.windowscommunicationsapps_8wekyb3d8bbwe\LocalState\Indexed\LiveComm \6e4f9dff0b76dd9b\1207120049\People\AddressBook\26000001_bef42d234ebd42.appcontent-ms*. Each contact formed an opening and closing 'properties' tag to house the search properties such as search keywords, full names, home addresses, birth dates, phone numbers, and other information as detailed in Fig 19, which may be of value for user profiling. Additionally, the similar information could be located for the user in the *%Appdata%\Local\Packages\ microsoft.windowscommunicationsapps_8wekyb3d8bbwe\LocalState\Indexed\LiveComm \6e4f9dff0b76dd9b\120712-0049\People\Me\24000001_*7b20c4c2b2382.appcontent-ms file.

## 5.4 IM Conversations and Transferred Files

Examinations of the directory listings determined that the files downloaded were saved in *% Downloads%\Microsoft.SkypeApp_kzf8qxf38zg5c!App\* and *%AppData%\Local\Packages\Microsoft.SkypeApp_kzf8qxf38zg5c\LocalState\<Skype name>\ReceiveStorage\* by default; each of which was given an ADS ZoneID with reading 'ZoneID = 3'. Meanwhile, copies of the transferred files were located in *%AppData%\Local\Packages\Microsoft.SkypeApp_kzf8qxf38zg5c\LocalState \<Skype name>\SendingStorage\*. The files retained the original filenames and extensions. In addition to the file download or transfer directory paths, we were able to recover copies of thumbnail images for the transferred or downloaded files within the Windows thumbcache.

An inspection of the registry entries observed that each transferred or downloaded file created a Globally Unique Identifier (GUID) key in *HKEY_USERS\<SID>\Software\Classes \LocalSettings\Software\Microsoft\Windows\CurrentVersion\AppModel\SystemAppData \Microsoft.SkypeApp_kzf8qxf38zg5c\PersistedStorageItemTable\ManagedByApp\*. The entries

```
<?xml version="1.0" ?>
- <config version="1.0" serial="78" timestamp="1421686251.63">
...
<Account>
  <LastPartnerId>830</LastPartnerId>
  <LastUsed>1421679670</LastUsed>
  <LocalData>4215</LocalData>
  <Migration>63</Migration>
  <OMigration>1</OMigration>
  </Account>
...
- <u>
  <echo123>9db4df93:2</echo123>
  <harold.2Ecornwall1>4857bb98:2</harold.2Ecornwall1>
  </u>
...
```

**Fig 18. Portion of config.xml file.**

doi:10.1371/journal.pone.0150300.g018

```
<?xml version="1.0" encoding="utf-16" ?>
<ContactAppData>
  <Properties
xmlns="http://schemas.microsoft.com/Search/2013/ApplicationContent">
    <Name>Harold Cornwall</Name>
    <Keywords>
      <Keyword>Harold</Keyword>
      <Keyword>Cornwall</Keyword>
      <Keyword>harold.cornwall1</Keyword>
      <Keyword>Harold Cornwall</Keyword>
    </Keywords>
    <AdditionalProperties>
      <Property Key="System.ItemType">
        <Value>Search.Contact</Value>
      </Property>
      <Property Key="System.Kind">
        <Value>Contact</Value>
      </Property>
      <Property Key="System.Contact.FirstName">
        <Value>Harold</Value>
      </Property>
      <Property Key="System.Contact.LastName">
        <Value>Cornwall</Value>
      </Property>
      <Property Key="System.AppUserModel.PackageRelativeApplicationID">
        <Value>Microsoft.WindowsLive.People</Value>
      </Property>
```

```
Continued..
      <Property Key="System.AppUserModel.ActivationContext">
        <Value>26000001</Value>
      </Property>
      <Property Key="System.Contact.HomeAddress1Locality">
        <Value>Malacca</Value>
      </Property>
      <Property Key="System.Contact.HomeAddress1Country">
        <Value>my</Value>
      </Property>
      <Property Key="System.Contact.Birthday">
        <Value>1990-02-02</Value>
      </Property>
      <Property Key="System.Contact.DisplayMobilePhoneNumbers">
        <Value>          </Value>
      </Property>
      <Property Key="System.Contact.ConnectedServiceIdentities">
        <Value>skype.com:harold.cornwall1</Value>
      </Property>
      <Property Key="System.Contact.DataSuppliers">
        <Value>skype.com</Value>
      </Property>
      <Property Key="System.Contact.PhoneNumbersCanonical">
        <Value>          6</Value>
      </Property>
    </AdditionalProperties>
  </Properties>
</ContactAppData>
```

Fig 19. An excerpt of the .APPCONTENT-MS file recovered in our research.

doi:10.1371/journal.pone.0150300.g019

of particular interest with the key are 'FilePath' and 'LastUpdatedTime', which hold the directory path and last modified time for the file. When the sample files were opened, references were found for the directory paths and last accessed times in the 'RecentDocs' registry key and 'DLLHOST.EXE.pf' prefetch file.

An inspection of the main.db database located further details regarding the file transfer or download in the 'Transfers' table. The details included the senders' names, transfer types (where 1 indicates receiving and 2 indicates transferring), reasons for transfer failure (if any), storage paths, the times when the transfers were accepted, started and finished, as well as other file transfer information as shown in Fig 20. Records specific to the conversation or file transfer threads were located in the 'Messages' table, which encompassed the senders' Skype names (authors), whether the correspondents were the user's permanent contacts, the times when the threads were sent in Unix epoch format, the message sending status and types (as indicated in Table 6), reasons for message sending failure (if any), and other information as shown in Fig 21. The group chat could be discerned from the 'participant_count' table column given the value higher than 2. Moreover, it was also possible to recover the conversation texts and

| | id | is_permanent | type | partner_handle | partner_dispname | status | failurereason | starttime | finishtime | filepath | filename | filesize | bytestransferred | bytespersecond | |
|---|---|---|---|---|---|---|---|---|---|---|---|---|---|---|---|
| | Fil... | Filter | Filter | Filter | Filter | Filter | Filter | Filter | Filter | Filter | Filter | Filter | Filter | Filter | Fi |
| 1 | 53 | 1 | 2 | adam.thomson11 | Adam Thomson | 2 | NULL | 1421685822 | 0 | C:\Users\anonymous\AppData\Local\Package... | SuspectToVictim.docx | 78080 | 0 | 0 | + |
| 2 | 54 | 1 | 2 | adam.thomson11 | Adam Thomson | 2 | NULL | 1421685822 | 0 | C:\Users\anonymous\AppData\Local\Package... | SuspectToVictim.jpg | 28937 | 0 | 0 | + |
| 3 | 55 | 1 | 2 | adam.thomson11 | Adam Thomson | 2 | NULL | 1421685822 | 0 | C:\Users\anonymous\AppData\Local\Package... | SuspectToVictim.pdf | 31747 | 0 | 0 | + |
| 4 | 56 | 1 | 2 | adam.thomson11 | Adam Thomson | 2 | NULL | 1421685822 | 0 | C:\Users\anonymous\AppData\Local\Package... | SuspectToVictim.rtf | 43360 | 0 | 0 | + |
| 5 | 57 | 1 | 2 | adam.thomson11 | Adam Thomson | 2 | NULL | 1421685822 | 0 | C:\Users\anonymous\AppData\Local\Package... | SuspectToVictim.txt | 2734 | 0 | 0 | + |
| 6 | 58 | 1 | 2 | adam.thomson11 | Adam Thomson | 2 | NULL | 1421685822 | 0 | C:\Users\anonymous\AppData\Local\Package... | SuspectToVictim.zip | 30967 | 0 | 0 | + |

Fig 20. The 'Transfers' table of main.db database.

doi:10.1371/journal.pone.0150300.g020

Table 6. Details of the 'Messages' table [66].

| Table field | Value | Description |
|---|---|---|
| **Type** | 4 | Conference |
| | 30/39 | Video session started/ended |
| | 50/51 | Contact ask/permission |
| | 53 | Blocked |
| | 60 | Sent emoticon |
| | 61 | Sent text message |
| | 63 | Sent contact details |
| | 64 | Sent SMS |
| | 67 | Sent voice message |
| | 68 | Sent file |
| | 110 | Date of birth |
| **chatmsg_type** | 3 | Text message |
| | 5 | Group chat |
| | 7 | Data transfer |
| | 18 | Added contact |
| **chatmsg_status** | 2 | Text message sent |
| | 4 | Text message downloaded |

doi:10.1371/journal.pone.0150300.t006

metadata associated with the downloaded or transferred files in the 'body_xml' table column (of the 'Messages' table). As can be seen in Fig 22, each downloaded or transferred file forms an opening and closing XML subtag (in the 'files' tag) to record its file size, transfer index, transfer ID, and filename in the 'body_xml' table column.

Another file of forensic interest that will potentially allow a practitioner to recover the conversation history is the 'Chatsync' file located in *%AppData%\Local\Packages\Microsoft.SkypeApp_kzf8qxf38zg5c\LocalState\<Skype name>\Chatsync\*. The 'Chatsync' file is stored in the format of <Random sixteen character strings>.DAT and is mainly used to facilitate chat log synchronisation between devices [67]. The 'Chatsync' file is chat-session-specific in the sense that a chatsync file is generally created for each chat session. Fig 23 illustrates that the 'Chatsync' files may provide the conversation texts and timestamp information for the chat sessions associated with the Skype user.

Fig 21. The 'Messages' table of main.db database.

doi:10.1371/journal.pone.0150300.g021

```
<files alt="">
    <file size="78080" index="0" tid="1335338368">SuspectToVictim.docx</file>
    <file size="287937" index="1" tid="358042097">SuspectToVictim.jpg</file>
    <file size="31747" index="2" tid="3482891630">SuspectToVictim.pdf</file>
    <file size="43360" index="3" tid="3018727815">SuspectToVictim.rtf</file>
    <file size="2734" index="4" tid="1324086924">SuspectToVictim.txt</file>
    <file size="30967" index="5" tid="621137037">SuspectToVictim.zip</file>
</files>
```

**Fig 22. File transfer metadata recovered from the 'body_xml' table column of the 'Messages' table.**

doi:10.1371/journal.pone.0150300.g022

Unsurprisingly, a manual search for terms unique to the Enron sample files (i.e., 'pensive' and 'parakeet') as well as table column names of the main.db database produced matches to the plain text copies of the transferred/downloaded files and main.db database in the unstructured datasets, respectively. However, there was no common footer information that could enable future carving of the main.db database. We also located fragments of the payloads for the conversation threads in the memory dump, which held the conversation times, senders and receivers' Skype names, and conversation texts as highlighted in Fig 24. When file transfers occurred, additional entries were observed for the filenames, file sizes, and file transfer IDs in the payload. The header fields could be suitable search terms for the remnants; a Yarascan search would attribute the remnants to the Skype's process.

Examination of the network traffic observed that the host established a direct UDP connection with the correspondents during conversations and file transfers, and hence the IP addresses could be detected. However, there was no definitive port number or URL which could enable future identification of the traffic. Further analysis of the network packets determined that the data were fully encrypted, but we were able to estimate when the conversations were taken place from the corresponding timestamp information.

## 5.5 Voice and Video Calls

Skype allows users to perform voice calls via the free Skype to Skype calls and in the premium version, users could make Skype to mobile or landline calls using Skype credit. In order to enhance the user's interactive experience, Skype allows users to share free video calls with anyone who has Skype and a webcam or compatible smartphone.

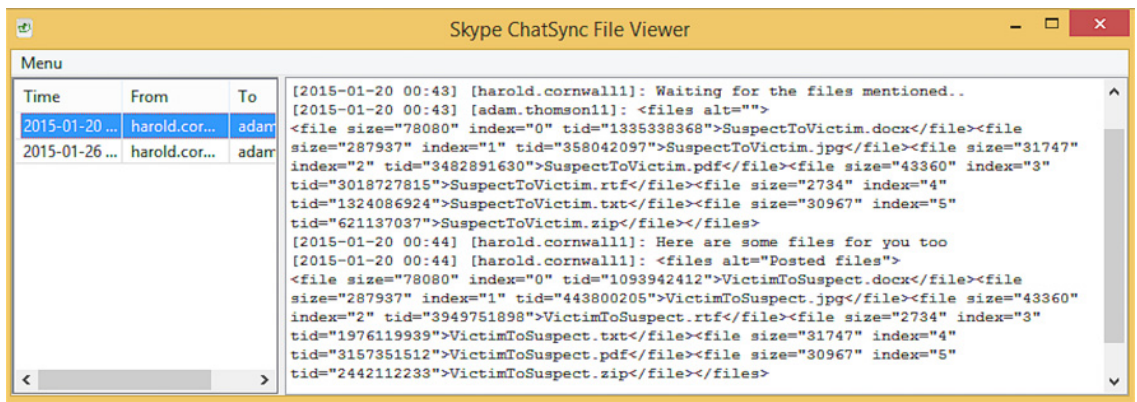

**Fig 23. Portion of the output from Skype Chatsync Reader.**

doi:10.1371/journal.pone.0150300.g023

**Fig 24. Remnants of Skype's payload header recovered from RAM.**

doi:10.1371/journal.pone.0150300.g024

Examinations of the directory listings determined that the Skype app does not save the voice and video calls. However, we were able to recover a wealth of caches relating to the voice and video calls in the main.db database. Recalling the 'Messages' table, it was observed that entries of the voice or video calls could be differentiated from the 'type' table column given the value 30, 39, or 67 (see Table 6). Details of the voice or video calls were recovered from the 'Calls' table, which comprised the callers' Skype names, the times when the calls were started, the call durations in seconds, and whether the calls were incoming calls, conference calls, and put on hold—see Fig 25. Additionally, the 'CallMembers' table provided additional information associated with the contacts with whom the user had voice or video calls such as the Skype names, full names, call charges, reasons for call failures (if any), graphical user IDs (represented in '<User's Skype name>-<Correspondent's Skype name>-<Call name>'), external IP addresses of the correspondents, call statuses, the times when the calls were started, the call durations, whether the calls were incoming or outgoing, conference calls, and from permanent contacts.

Examinations of the network traffic of the voice and video calls observed that the app established a session with the CloudFlare (GlobalSign) server for Online Certificate Status Protocol (OSCP) stapling and with the Verisign server for certificate authentication. When the calls occurred, the IP addresses were allocated to the supernodes (on random TCP ports) and then to the Windows Live server (i.e., IP address 65.55.246.85) on port 443, which we theorised for user lookups and authentications. The network traffic was subsequently seen with random IP addresses and UDP ports, which were hypothesised from supernodes responsible for bridging the VoIP, but the contents were encrypted completely.

| id | is_permanent | begin_timestamp | topic | mut | is_unseen_missed | host_identity | e_sti | duration | indle | ss_te | active_members | is_active | name | video_disabled | d_exi | r_ide | put | is_incoming | is_conference |
|---|---|---|---|---|---|---|---|---|---|---|---|---|---|---|---|---|---|---|---|
| | Fil... | Filter | Filter | Fil... | Fil... | Filter | Filter | Fil... | Filter | Fil... | N... | Filter | Filter | Filter | N... | N... | N... | Filter | Filter |
| 1 | 100 | 1 | 1421685999 | | 0 | 0 | adam.thomson11 | 0 | 14 | 0 | N... | 0 | | 0 | 8-1421685999 | NULL | N... | N... | N... | 1 | NULL | N... |
| 2 | 104 | 1 | 1421686068 | N... | N... | 0 | adam.thomson11 | 0 | 0 | 0 | N... | 0 | | 0 | 9-1421686068 | NULL | N... | N... | N... | 0 | NULL | N... |
| 3 | 110 | 1 | 1421686088 | | 0 | 0 | adam.thomson11 | 0 | 11 | 0 | N... | 0 | | 0 | 10-1421686088 | NULL | N... | N... | N... | 0 | NULL | N... |
| 4 | 116 | 1 | 1421686142 | N... | 0 | 0 | adam.thomson11 | 0 | 0 | 0 | N... | 0 | | 0 | 11-1421686142 | NULL | N... | N... | N... | 0 | NULL | N... |

**Fig 25. An excerpt of the 'Calls' table of main.db database.**

doi:10.1371/journal.pone.0150300.g025

```
<videomessage sid="90699566cef64bd97b99704588c41609" feature_name="" publiclink="https://vm.skype.com/mail/adam.thomson11/90699566cef64bd97b99704588c41609">You've got a
video message. To see it, simply copy this secret code 1400 and <a href='https://vm.skype.com/mail/adam.thomson11/90699566cef64bd97b99704588c41609'>watch your video message
here</a>
</videomessage>
```

**Fig 26. Video message metadata recovered from the 'body_xml' table column of the 'Messages' table.**

## 5.6 Video Messages

Skype allows the users to share video messages (video recordings) with other online and offline users. The video messages are sent as a link in Skype version 6.5 or older, which requires a secret code access.

Sending a video message, it was observed that the Skype app stored a copy of the video message in *%AppData%\Local\Packages\Microsoft.SkypeApp_kzf8qxf38zg5c\LocalState\<Skype name>\media\* of the sender's device by default. The video message also created a thumbnail image in *%AppData%\Local\Packages\Microsoft.SkypeApp_kzf8qxf38zg5c\LocalState\<Skype name>\thumbnails\.*

Analysis of the main.db database revealed that the Skype app cached notifications of the video messages in the 'body_xml' table column of the 'Messages' table, and the entry of which could be discerned from the XML tag 'videomessage'. The notification records provided the video message IDs, public links, and secret codes (sent from Skype application version 6.5 or older) for the video messages sent or received by the user as highlighted in Fig 26. Meanwhile, details of the video messages sent/received could be located in the 'VideoMessages' table, which included the directory paths, public links, titles, descriptions (if any), author names, creation times, transferring or receiving times as illustrated in Fig 27.

## 5.7 Uninstallation of the Skype App

Uninstallation of the Skype app did not remove the installation folders like as was presented for the Facebook app. However, the application folder was removed from the file system completely. Analysis of the unallocated space, RAM, as well as a variety Windows system files (i.e., $LogFile, $MFT, $UsnJrnl, pagefile.sys, shortcuts, event logs, prefetch files, and thumbcache files) resulted in the recovery of artefacts created prior to uninstallation of the app, with additional references to the directory paths and timestamp information for the files removed during the uninstallation in $LogFile, $MFT, $UsnJrnl.

## 6. Discussion

In this research, we identified artefacts common to investigating the Windows Store apps for IM. Previous studies only addressed dead analysis of the IM apps, while we focus on both the volatile and non-volatile artefacts. Our experiments showed that the Facebook and Skype apps maintain a wealth of caches of forensic interest within the 'localstate' application folder in Sqlite database unencrypted, which seem to agree with the findings of Lee and Chung [34]. This indicated that when a user has used a Windows Store app for IM, there will be records remaining in the application folder to support reconstruction of the logins, contact lists, conversations, file transfers, and other relevant IM activities, assuming that the app is not removed.

| id | is_permanent | qik_id | attached_msg_ids | sharing_id | status | vod_status | vod_path | local_path | public_link | progress | title | description | author | reation_timestam | type |
|---|---|---|---|---|---|---|---|---|---|---|---|---|---|---|---|
| Fil... | Filter | Filter | Filter | Filter | Filter | Filter | Filter | Filter | Filter | Filter | Fil... | Filter | Filter | Filter | Filter |
| 1 | 173 | 1 | ✏✐⌀... | 5577 | 90699566cef... | 3 | 3 | C:\Users\anonym... | C:\Users\anonym... | https://vm.skype.com/... | 100 | | | adam.thomson11 | 1422254107 | |

**Fig 27. An excerpt of the 'VideoMessages' table of main.db database.**

Although several registry keys new to the Windows Store apps could be recovered, it was determined that the Windows Store apps record significantly less information of interest to IM forensics in comparison to traditional client desktop application. While artefacts of the user profiles, contact lists and recent communications could be potentially recovered from the registry of the older Windows IM client applications [16, 21, 36–38, 42, 43], only installation metadata (i.e., install paths and times) could be recovered for the Windows Store apps, albeit records of the transferred files could be recovered in some cases. This is likely resulted from the adoption of the app caches. Similar to any other Windows client applications, our examinations of the system files such as $LogFile, $MFT, $UsnJrnl, shortcuts, event logs, thumbnail cache, as well as the 'recentdocs' registry key revealed that additional timestamp information could be recovered to support evidence found in all scenarios, but results may not be definitive.

It should be noted, however, that that the significance, amount, and location of artefacts could vary in accordance to the Windows Store apps under investigation. For instance, in our research, it was determined that:

- both the Facebook and Skype apps maintain a different directory structure in the application folders;

- the apps hold different database schema for the application caches;

- caches of the Facebook app appear only when the user is logged in from the app, while caches of the Skype app remain resident throughout the lifetime of the app;

- the Skype app caches copies of the transferred and downloaded files in the application folder but this is not the case with the Facebook app;

- only the Skype app holds records of the transferred or downloaded files in *HKEY_USERS \<SID>\Software\Classes\LocalSettings\Software\Microsoft\Windows\CurrentVersion\App-Model\SystemAppData\<Package ID>\PersistedStorageItemTable\ManagedByApp\*.

The findings suggested that while a method can be generally defined to guide the investigation of the Windows Store apps, a different process may be necessary for investigating the different IM apps.

Our examinations of the physical memory captures indicated that the memory dumps can provide a potential alternative method for recovery of the application caches in plain text, with the exception of the login password. The fact that there was no clear text password in the hard drives and memory dumps should perhaps be unsurprising since the credential information is securely encrypted in the Credential Locker [29]. Nevertheless, a practitioner must keep in mind that memory changes frequently according to users' activities and will be wiped as soon as the system is shut down.

In some cases, remnants of the caches could be located in the swap file (pagefile.sys) and unallocated space. The most likely explanation for the remnants is that the system swapped inactive memory pages containing the application caches out of the memory to the hard disk during the system's normal operation. As the remnants were recovered with minimal space configuration in our research, we believe there will be a greater chance of remnants on a typically larger system. Although the network traffic was encrypted, sufficient IP address and URL references could be located for scoping the user activities as well as requesting for assistance from counterparts overseas (i.e., via Interpol). Hence, we recommend that the physical memory and network captures should be undertaken wherever practical. Table 7 summarises the key artefacts located as part of our research.

Table 7. Summary of findings.

| Source of evidence | Facebook app | Skype app |
|---|---|---|
| **Registry branches of forensic interest.** | *HKEY_USERS\<SID>\Software\Classes\Local Settings \Software\Microsoft\Windows\CurrentVersion\AppModel \Repository\Families\Faceook.Facebook_8xx8rvfyw5nnt \Facebook.Facebook_1.4.0.9_x64_8xx8rvfyw5nnt* | *HKEY_USERS \<SID>\Software\Classes\Local Settings\Software \Microsoft\Windows\CurrentVersion\AppModel\Repository\Families \Microsoft.SkypeApp_kzf8qxf38zg5c\Microsoft. SkypeApp_2.0.0.5011_x86__kzf8qxf38zg5c* |
| | *Software\Microsoft\Windows\CurrentVersion\Explorer \RecentDocs* | *HKEY_USERS\<SID>\Software\Classes\Local Settings\Software \Microsoft\Windows\CurrentVersion\AppModel\SystemAppData \Microsoft.SkypeApp_kzf8qxf38zg5c\PersistedStorageItemTable \ManagedByApp\<GUID>* |
| | | *Software\Microsoft\Windows\CurrentVersion\Explorer\RecentDocs* |
| **Directory paths/ files of forensic interest** | *%AppData%\Local\Temp\winstore.log* | *%AppData%\Local\Temp\winstore.log* |
| | *%AppData%\Local\Packages\winstore_cw5n1h2txyewy\AC \Temp\winstore.log* | *%AppData%\Local\Packages\winstore_cw5n1h2txyewy\AC\Temp \winstore.log* |
| | Analytics.sqlite, FriendRequest.sqlite, Friends. sqlite, Messages.sqlite, Notifications.sqlite, and Stories. sqlite databases stored in *%AppData%\Local\Packages \Facebook.Facebook_8xx8rvfyw5nnt\LocalState\<User-specific Facebook ID>\DB\* | User-specific *%AppData%\Local\Packages\Microsoft. SkypeApp_kzf8qxf38zg5c\LocalState\<Skype name>\main.db* |
| | Caches of the downloaded files stored in *%AppData%\Local \Packages\Package ID\AC\INetCache\Cache ID\* | *%AppData%\Local\Packages\Microsoft.SkypeApp_kzf8qxf38zg5c \LocalState\shared.xml* |
| | | *%AppData%\Local\Packages\Microsoft.SkypeApp_kzf8qxf38zg5c \LocalState\<Skype name>\Chatsync\* |
| | | *%AppData%\Local\Packages\Microsoft.SkypeApp_kzf8qxf38zg5c \LocalState\avatars\* |
| | | *%Downloads%\Microsoft.SkypeApp_kzf8qxf38zg5c\App\* |
| | | *%AppData%\Local\Packages\Microsoft.SkypeApp_kzf8qxf38zg5c \LocalState\<Skype name>\ReceiveStorage\* |
| | | *%AppData%\Local\Packages\Microsoft.SkypeApp_kzf8qxf38zg5c \LocalState\<Skype name>\SendingStorage\* |
| | | APPCONTENT-MS files located in *%Appdata%\Local\Packages \microsoft.windowscommunicationsapps_8wekyb3d8bbwe \LocalState\Indexed\LiveComm\6e4f9dff0b76dd9b\120712--0049 \People\AddressBook\* and *%Appdata%\Local\Packages\microsoft. windowscommunicationsapps_8wekyb3d8bbwe\LocalState \Indexed\LiveComm\6e4f9dff0b76dd9b\120712--0049\People\Me\* |
| **Prefetch files** | FACEBOOK.EXE.pf | WWAHOST.EXE.pf |
| | | DLLHOST.EXE.pf |
| **Link files** | Located link files for the transferred or downloaded files in *% \AppData\Roaming \Microsoft\Windows\Recent\* | Located link files for the login page as well as the transferred or downloaded files in *%\AppData\Roaming \Microsoft\Windows \Recent\* |
| **Thumbcache files** | Thumbnail images for the transferred or downloaded files | Thumbnail images for the transferred or downloaded files |
| | Profile pictures of the user and the contacts | Avatars of the user and the contacts |
| **Swap files and physical memory dumps** | Copies of the files of forensic interest as well as transferred or downloaded files unencrypted | Copies of the files of forensic interest as well as transferred or downloaded files in plain text |
| | Filename and path references for the files of forensic interest and transferred or downloaded files | Filename and path references for the files of forensic interest and transferred or downloaded files |
| | The process name could be discerned from 'Facebook.exe' | Payload headers for the IM and file transfer threads |
| | | The process name could be discerned from 'WWAHost.exe' |
| **Unallocated space** | Copies of the files of forensic interest as well as transferred or downloaded file in plain text | Copies of the files of forensic interest as well as transferred or downloaded file in plain text |

(Continued)

**Table 7.** (Continued)

| Source of evidence | Facebook app | Skype app |
|---|---|---|
| | Filename and path references for the files of forensic interest and transferred or downloaded files | Filename and path references for the files of forensic interest and transferred or downloaded files |
| Network traffic | Host and servers' IP addresses | Host and servers' IP addresses |
| | Associated timestamps | Host and correspondents' IP addresses |
| | Web documents and image files from the HTTP sites. | Associated timestamps |
| | | Web documents and image files from the HTTP sites. |

doi:10.1371/journal.pone.0150300.t007

## 7. Conclusion and Future Work

Instant messaging (IM), such as VoIP apps, are increasingly popular among individuals and business organisations [68], including criminals. To ensure the most effective collection of evidence of relevance, it is important that a practitioner possess an up-to-date understanding of different technologies [69–77]. This paper presented the findings from our forensic examination (acquisition and reconstruction of the terrestrial artefacts left by the use) of two popular Windows Store IM apps, namely Facebook and Skype. The study consisted of installation, uninstallation, logins, conversations, transferred files, and and other IM activities specific to the apps investigated.

The results indicated that use of the Windows Store apps IM apps can leave behind incriminating evidential material useful or critical to an investigation on the hard drive, memory dumps, and network captures. The artefacts located as part of our experiments are likely to be common with other Windows Store IM apps as well as newer Windows OS (i.e., Windows 10), since the apps share a common feature set. While the implementation may vary between different IM apps, we contended that practitioners could use the artefacts identified in this research as a basis for their investigation of the client as a potential evidence source.

Future work would include:

1. Extending this study to new (version of) apps, including apps popular in other countries (i.e., WeChat and LINE), to have an up-to-date forensic understanding of these technologies that can be used to inform investigations.

2. Proposing a method for analyzing new (as of yet) unknown apps with similar functionality (ies). If such a method can be developed, evaluation might demonstrate that it can it be applied to a new app, or even implemented into a tool.

## Author Contributions

Conceived and designed the experiments: TYY AD KKRC. Performed the experiments: TYY. Analyzed the data: TYY. Contributed reagents/materials/analysis tools: TYY AD KKRC. Wrote the paper: TYY AD KKRC ZM.

## References

1. The Radicati Group Releases "Instant Messaging Statistics Report, 2015–2019. California: Radicati Group; 2015 March 16. Available: http://www.radicati.com/?p=13001. Accessed 18 June 2015.

2. Online dating fraud up by 33% last year. London: City of London Police; 2015 [2015 February 13] Available: https://www.cityoflondon.police.uk/advice-and-support/fraud-and-economic-crime/nfib/nfib-news/Pages/online-dating-fraud.aspx. Accessed 29 May 2015

3. Meyers SL. Special Report, Part 1: "Diploma mill" scams continue to plague Milwaukee's adult students. Washington: Milwaukee Neighborhood News Service; 2014 May 21. Available: http://

milwaukeenns.org/2014/05/21/special-report-diploma-mill-scams-continue-to-plague-milwaukees-adult-students/. Accessed 24 May 2015

4. Timoney N. Consumer Contact: Job Advertising Fraud. Bangor: WABI TV5; 2014 May 12. Available: http://wabi.tv/2014/05/12/consumer-contact-job-advertising-fraud/. Accessed 24 May 2015

5. Instant messaging Trojan spreads through the UK. [Place unknown]: Help Net Security. 2014 May 27. Available: http://www.net-security.org/malware_news.php?id=2773. Accessed 24 May 2015

6. Barnes T. Margate pedophile jailed for five years. U.K: Thanet Gazette; 2014 April 7. Available: http://www.thanetgazette.co.uk/Margate-paedophile-jailed-years/story-20922860-detail/story.html. Accessed 24 May 2015

7. Godfrey M. Pedophiles coercing kids using phone app. Sydney: Sydney Morning Herald; 2014. Available: http://news.smh.com.au/breaking-news-national/pedophiles-coercing-kids-using-phone-app-20130327-2gu3a.html. Accessed 24 May 2015

8. McCallum N. Pedophile posed as Bieber to lure victims. Australia: Mi9;2013. Available: http://www.9news.com.au/world/2013/09/17/10/30/pedophile-posed-as-bieber-to-lure-victims. Accessed 24 May 2015

9. Jacksonville Man Sentenced in Child Pornography Case. Raleigh: The Federation Bureau of Investigation (FBI); 2015. Available: http://www.fbi.gov/charlotte/press-releases/2015/jacksonville-man-sentenced-in-child-pornography-case. Accessed 20 May 2015.

10. Norouzizadeh Dezfouli F, Dehghantanha A, Eterovic-Soric B, Choo K-KR. Investigating Social Networking applications on smartphones detecting Facebook, Twitter, LinkedIn and Google+ artefacts on Android and iOS platforms. Australian Journal of Forensic Sciences. 2015 Aug 7;1–20.

11. Ali D. Mining the Social Web: Data Mining Facebook, Twitter, LinkedIn, Google+, Github, and More. Journal of Information Privacy and Security. 2015 Apr 3; 11(2):137–8.

12. Investigative Uses of Technology: Devices, Tools, and Techniques. U.S: National Criminal Justice Reference Service (NCJRS); 2007 October 3. Available: https://www.ncjrs.gov/pdffiles1/nij/213030.pdf. Accessed 4 May 2015.

13. Barghuthi NBA, Said H. Social Networks IM Forensics: Encryption Analysis. Journal of Communications. 2013; 8: 708–715. doi: 10.12720/jcm.8.11.708–715

14. Golden TW, Skalak SL, Clayton MM. A Guide to Forensic Accounting Investigation. 2 edition. Hoboken, N.J: Wiley; 2011.

15. Procure Secure: A guide to monitoring of security service levels in cloud contracts—ENISA. Europe: European Union Agency for Network and Information Security (ENISA); 2012 April 2. Available: https://www.enisa.europa.eu/activities/Resilience-and-CIIP/cloud-computing/procure-secure-a-guide-to-monitoring-of-security-service-levels-in-cloud-contracts. Accessed 10 December 2015

16. Dickson M. An examination into AOL Instant Messenger 5.5 contact identification. Digital Investigation. 2006; 3: 227–237. doi: 10.1016/j.diin.2006.10.004

17. Martini B, Choo K-KR. An integrated conceptual digital forensic framework for cloud computing. Digital Investigation. 2012; 9: 71–80. doi: 10.1016/j.diin.2012.07.001

18. Quick D, Martini B, Choo R. Cloud Storage Forensics. Syngress; 2013.

19. Kiley M, Dankner S, Rogers M. Forensic Analysis of Volatile Instant Messaging. In: Ray I, Shenoi S, editors. Advances in Digital Forensics IV. Springer US; 2008. p. 129–38. Available: http://link.springer.com/chapter/10.1007/978-0-387-84927-0_11. Accessed 11 June 2015.

20. Forensic Investigation of Instant Messenger Histories. [Place unknown]: Forensic Focus; [Date unknown]. Available: http://www.forensicfocus.com/forensic-investigation-of-instant-messenger-histories. Accessed 24 May 2015.

21. Reust J. Case study: AOL instant messenger trace evidence. Digital Investigation. 2006; 3: 238–243. doi: 10.1016/j.diin.2006.10.009

22. Carvey H. Instant messaging investigations on a live Windows XP system. Digital Investigation. 2004 Dec; 1(4):256–60.

23. Quick D, Choo K-KR. Dropbox analysis: Data remnants on user machines. Digital Investigation. 2013; 10: 3–18. doi: 10.1016/j.diin.2013.02.003

24. Quick D, Choo K-KR. Google Drive: Forensic Analysis of Data Remnants. Journal of Network Computing and Application. 2014; 40: 179–193. doi: 10.1016/j.jnca.2013.09.016

25. Quick D, Choo K-KR. Digital droplets: Microsoft SkyDrive forensic data remnants. Future Generation Computer Systems. 2013; 29: 1378–1394. doi: 10.1016/j.future.2013.02.001

26. Brockschmidt K. Programming Windows Store Apps with HTML, CSS, and JavaScript. Microsoft Press; 2014

27. Mehreen S, Aslam B. Windows 8 cloud storage analysis: Dropbox forensics. In IEEE; 2015. p. 312–7. Available: http://ieeexplore.ieee.org/lpdocs/epic03/wrapper.htm?arnumber=7058522. Accessed 6 April 2015.

28. Fleming R. How many devices can you install a Windows 8 app on?. U.S: Microsoft Corporation; 2013 October 1. Available: http://blogs.msdn.com/b/education/archive/2013/10/01/how-many-devices-can-you-install-a-windows-8-app-on.aspx. Accessed 28 March 2015

29. How to store user credentials (XAML). U.S: Microsoft; [Date unknown]. Available: https://msdn.microsoft.com/en-us/library/windows/apps/xaml/Hh465069(v=win.10).aspx. Accessed 24 May 2015.

30. Sanna P, Wright A. Windows 8.1 Absolute Beginner's Guide. Que Publishing; 2013.

31. Thomson A. Windows 8 Forensic Guide. Washington; The George Washington University; 2012. Available: http://propellerheadforensics.files.wordpress.com/2012/05/thomson_windows-8-forensic-guide2.pdf. Accessed 13 May 2015.

32. Rasmussen B, High-Performance Windows Store Apps. Microsoft Press; 2014.

33. Iqbal A, Al Obaidli H, Marrington A, Jones A. Windows Surface RT tablet forensics. Digital Investigation. 2014 May; 11, Supplement 1: S87–S93.

34. Lee C, Chung M. Digital Forensic Analysis on Window8 Style UI Instant Messenger Applications. In: Park JJ (Jong H, Stojmenovic I, Jeong HY, Yi G, editors. Computer Science and its Applications. Springer Berlin Heidelberg; 2015. p. 1037–42. Available: http://link.springer.com/chapter/10.1007/978-3-662-45402-2_147. Accessed 22 March 2015.

35. Carvey H. Windows Forensic Analysis Toolkit: Advanced Analysis Techniques for Windows 8. Elsevier; 2014.

36. Dickson M. An examination into MSN Messenger 7.5 contact identification. Digital Investigation. 2006 Jun; 3(2):79–83.

37. Dickson M. An examination into Yahoo Messenger 7.0 contact identification. Digital Investigation. 2006 Sep; 3(3):159–65

38. Dickson M. An examination into Trillian basic 3.x contact identification. Digital Investigation. 2007 Mar; 4(1):36–45.

39. Yasin M, Abulaish M. DigLA—A Digsby log analysis tool to identify forensic artifacts. Digital Investigation. 2013 Feb; 9(3–4):222–34.

40. Yasin M, Kausar F, Aleisa E, Kim J. Correlating messages from multiple IM networks to identify digital forensic artifacts. Electron Commer Res. 2014 Sep 18; 14(3):369–87

41. Yasin M, Abulaish M, Elmogy MNN. Forensic Analysis of Digsby Log Data to Trace Suspected User Activities. In: Park JH (James), Kim J, Zou D, Lee YS, editors. Information Technology Convergence, Secure and Trust Computing, and Data Management. Springer Netherlands; 2012. p. 119–26. Available: http://link.springer.com/chapter/10.1007/978-94-007-5083-8_16. Accessed 1 April 2015.

42. Van Dongen WS. Forensic artefacts left by Windows Live Messenger 8.0. Digital Investigation. 2007 Jun; 4(2):73–87.

43. Van Dongen WS. Forensic artefacts left by Pidgin Messenger 2.0. Digital Investigation. 2007 Sep; 4 (3–4):138–45.

44. Levendoski M, Datar T, Rogers M. Yahoo! Messenger Forensics on Windows Vista and Windows 7. In: Gladyshev P, Rogers MK, editors. Digital Forensics and Cyber Crime. Berlin, Heidelberg: Springer Berlin Heidelberg; 2012. p. 172–9. Available: http://link.springer.com/10.1007/978-3-642-35515-8_14. Accessed 6 April 2015.

45. Wong K, Lai ACT, Yeung JCK, Lee WL, Chan PH. Facebook Forensics. Singapore: Valkyrie-X Security Research Group; 2011 July. Available: www.fbiic.gov/public/2011/jul/Facebook_Forensics-Finalized.pdf. Accessed 12 May 2015.

46. Al Mutawa N, Al Awadhi I, Baggili I, Marrington A. Forensic artifacts of Facebook's instant messaging service. Internet Technology and Secured Transactions (ICITST), 2011 International Conference for. 2011. pp. 771–776.

47. Al Mutawa N, Baggili I, Marrington A. Forensic analysis of social networking applications on mobile devices. Digital Investigation. 2012 Aug; 9, Supplement: S24–S33.

48. Said H, Yousif A, Humaid H. IPhone forensics techniques and crime investigation. In IEEE; 2011. p. 120–5. Available: http://ieeexplore.ieee.org/lpdocs/epic03/wrapper.htm?arnumber=6107946. Accessed 4 July 2015.

49. Walnycky D, Baggili I, Marrington A, Moore J, Breitinger F. Network and device forensic analysis of Android social-messaging applications. Digital Investigation. 2015 Aug; 14, Supplement 1: S77–84.

50. Levinson A, Stackpole B, Johnson D. Third Party Application Forensics on Apple Mobile Devices. In: 2011 44th Hawaii International Conference on System Sciences (HICSS). 2011. p. 1–9.

51. Tso Y-C, Wang S-J, Huang C-T, Wang W-J. iPhone Social Networking for Evidence Investigations Using iTunes Forensics. In: Proceedings of the 6th International Conference on Ubiquitous Information Management and Communication. New York, NY, USA: ACM; 2012. p. 62:1–62:7. Available: http://doi.acm.org/10.1145/2184751.2184827. Accessed 8 December 2015.

52. Chu H-C, Deng D-J, Park JH. Live Data Mining Concerning Social Networking Forensics Based on a Facebook Session Through Aggregation of Social Data. IEEE Journal on Selected Areas in Communications. 2011 Aug; 29(7):1368–76.).

53. Wongyai W, Charoenwatana L. Examining the network traffic of facebook homepage retrieval: An end user perspective. 2012 International Joint Conference on Computer Science and Software Engineering (JCSSE). 2012. pp. 77–81. 10.1109/JCSSE.2012.6261929.

54. Sgaras C, Kechadi M-T, Le-Khac N-A. Forensics Acquisition and Analysis of Instant Messaging and VoIP Applications. In: Garain U, Shafait F, editors. Computational Forensics. Springer International Publishing; 2015. p. 188–99. Available: http://link.springer.com/chapter/10.1007/978-3-319-20125-2_16. Accessed 11 October 2015

55. Simon M, Slay J. Recovery of Skype Application Activity Data from Physical Memory. ARES '10 International Conference on Availability, Reliability, and Security, 2010. 2010. pp. 283–288. 10.1109/ARES.2010.73.

56. Teng S-Y, Lin Y-L. Skype Chat Data Forgery Detection. In: Kim T, Ko D, Vasilakos T, Stoica A, Abawajy J, editors. Computer Applications for Communication, Networking, and Digital Contents. Springer Berlin Heidelberg; 2012. pp. 108–114. Available: http://link.springer.com/chapter/10.1007/978-3-642-35594-3_15.

57. Baset SA, Schulzrinne HG. An Analysis of the Skype Peer-to-Peer Internet Telephony Protocol. INFO-COM 2006 25th IEEE International Conference on Computer Communications Proceedings. 2006. pp. 1–11. 10.1109/INFOCOM.2006.312.

58. Azab A, Watters P, Layton R. Characterising Network Traffic for Skype Forensics. Cybercrime and Trustworthy Computing Workshop (CTC), 2012 Third. 2012. pp. 19–27. 10.1109/CTC.2012.14.

59. McKemmish R. What is forensic computing? Canberra: Australian Institute of Criminology;1999 June. Available: http://www.aic.gov.au/media_library/publications/tandi_pdf/tandi118.pdf. Accessed 20 May 2015

60. Company Info. U.S: Facebook. [Date unknown]. Available: https://newsroom.fb.com/company-info/. Accessed 24 May 2015

61. Reisinger D. Windows 8.1 app updates: Facebook, Netfix, and more. U.S: CNET;2013 October 17. Available: http://www.cnet.com/news/windows-8-1-app-updates-facebook-netflix-and-more/. Accessed 4 May 2015

62. About URL Security Zones (Windows). U.S: Microsoft; [Date unknown]. Available: https://msdn.microsoft.com/en-us/library/ms537183.aspx#internet. Accessed 24 May 2015

63. Microsoft to Acquire Skype. U.S: Microsoft; 2011. Available: http://news.microsoft.com/2011/05/10/microsoft-to-acquire-skype/. Accessed 24 May 2015

64. Wurm K. Skype and a New Audio Codec. U.S: Skype; 2012 September 12. Available: http://blogs.skype.com/2012/09/12/skype-and-a-new-audio-codec/. Accessed 24 May 2015

65. Skype Forensics. U.S: InfoSec Institute; [Date unknown]. Available: http://resources.infosecinstitute.com/skype-forensics-2/.Accessed 24 May 2015.

66. Kuhlee L, Völzow V. Computer-Forensik Hacks. O'Reilly Germany; 2012.

67. McQuaid J. Skype Forensics: Analyzing Call and Chat Data From Computers and Mobile U.S: Magnet Forensics; 2012. Available: http://www.magnetforensics.com/wp-content/uploads/2014/04/Skype-Forensics-Analyzing-Call-and-Chat-Data-From-Computers-and-Mobile-Magnet-Forensics.pdf. Accessed 12 May 2015.

68. Azfar A, Choo K-KR, Liu L. Android mobile VoIP apps: A survey and examination of their security and privacy. Electronic Commerce Research. 2016. doi: 10.1007/s10660-015-9208-1

69. Azfar A, Choo K-KR, Liu L. An Android Social App Forensics Adversary Model. In Proceedings of Annual Hawaii International Conference on System Sciences (HICSS 2016). 2016. [In press].

70. Azfar A, Choo K-KR, Liu L. An Android Communication App Forensic Taxonomy. Journal of Forensic Sciences. 2016 [In press].

71. Azfar A, Choo K-KR, Liu L. Forensic Taxonomy of Popular Android mHealth Apps. In Proceedings of Americas Conference on Information Systems (AMCIS 2015). 2015. http://aisel.aisnet.org/cgi/viewcontent.cgi?article=1217&context=amcis2015.

72. Do Q, Martini B, Choo K-KR 2015. A Forensically Sound Adversary Model for Mobile Devices. PLOS ONE 10(9): e0138449. doi: 10.1371/journal.pone.0138449 PMID: 26393812

73. Farnden J, Martini B, Choo K-KR. Privacy Risks in Mobile Dating Apps. In Proceedings of Americas Conference on Information Systems (AMCIS 2015). 2015. http://aisel.aisnet.org/cgi/viewcontent.cgi?article=1427&context=amcis2015.

74. Immanuel F, Martini B, Choo K-KR. Android cache taxonomy and forensic process. In Proceedings of IEEE International Conference on Trust, Security and Privacy in Computing and Communications (TrustCom 2015). 2015: 1094–1101. 10.1109/Trustcom-BigDataSe-ISPA.2015.488.

75. Leom MD, D'Orazio C, Deegan G, Choo K-KR. Forensic Collection and Analysis of Thumbnails in Android. In Proceedings of IEEE International Conference on Trust, Security and Privacy in Computing and Communications (TrustCom 2015). 2015: 1059–1066. 10.1109/Trustcom-BigDataSe-ISPA.2015.483.

76. Ganji M, Dehghantanha A, Udzir NI, Damshenas M. Cyber warfare trends and future. Advances in Information Sciences and Service Sciences. 2013 Aug; 5(13): 1–10.

77. Mohtasebi S, Dehghantanha A, Broujerdi HG. Smartphone Forensics: A Case Study with Nokia E5-00 Mobile Phone. International Journal of Digital Information and Wireless Communications (IJDIWC). 2011; 1(3): 651–5.